\documentclass[aps,prd,preprint,superscriptaddress,nofootinbib]{revtex4-1}
\usepackage{braket}
\usepackage{xargs}
\usepackage{mathtools}
\usepackage[english]{babel}
\usepackage[utf8]{inputenc}
\usepackage{amsmath,amssymb,braket,slashed,mathrsfs}
\usepackage{pifont}
\usepackage{graphicx}
\usepackage{color,hyperref}
\usepackage{multirow,array}

\newcommand{\ba}{\begin{eqnarray}}
\newcommand{\ea}{\end{eqnarray}}

\newcommand{\innovation}{Collaborative Innovation Center of Quantum Matter, Beijing 100871, China}
\newcommand{\chep}{Center for High Energy Physics, Peking University, Beijing 100871, China}
\newcommand{\pkuphy}{School of Physics and State Key Laboratory of Nuclear Physics and Technology, Peking University, Beijing 100871,
China}

\newcommand{\Uconn}{Department of Physics, University of Connecticut, Storrs, CT 06269, USA}
\newcommand{\RBRC}{RIKEN-BNL Research Center, Brookhaven National Laboratory, Building 510, Upton, NY 11973}

\begin{document}
\title{Lattice QCD calculation of $K\to \ell\nu_\ell \ell'^+ \ell'^-$ decay width}

\author{Xin-Yu Tuo}\affiliation{\pkuphy}
\author{Xu~Feng}\email{xu.feng@pku.edu.cn}\affiliation{\pkuphy}\affiliation{\innovation}\affiliation{\chep}
\author{Lu-Chang Jin}
\affiliation{\Uconn}\affiliation{\RBRC}
\author{Teng Wang}\affiliation{\pkuphy}

\date{\today}

\begin{abstract}
	We develop a methodology for the computation of the $K\to \ell\nu_\ell \ell'^+ \ell'^-$ decay width
    using lattice QCD and present an exploratory study here.
    We use a scalar function method to account for the momentum
    dependence of the decay amplitude and adopt the  infinite volume reconstruction method to reduce the systematic
    errors such as the temporal truncation effects and the finite-volume effects. We then perform a four-body
    phase-space integral to obtain the decay width. The only remaining
    technical problem is the possible power-law finite-volume effects associated
    with the process of
    $K\to\pi\pi \ell\nu_\ell\to \ell\nu_\ell \ell'^+ \ell'^-$, where the
    intermediate state involves multiple hadrons. In this work, we use
    a gauge ensemble of twisted mass fermion with a pion mass $m_\pi=352$ MeV and a nearly-physical kaon mass.
    At this kinematics, the $\pi\pi$ in the intermediate state cannot be
    on shell simultaneously as $2m_\pi>m_K$ and the finite-volume effects associated with
    $\pi\pi$ state are exponentially suppressed.
    Using the developed methods mentioned above, we calculate the branching ratios 
    for four channels of $K\to \ell\nu_\ell\ell'^+ \ell'^-$, and
    obtain the results comparable to the experimental measurements and ChPT
    predictions. Our
    work demonstrates the capability of lattice QCD to improve Standard Model
    prediction in $K\to \ell\nu_\ell \ell'^+ \ell'^-$ decay width. 
\end{abstract}

\maketitle

\section{Introduction}
\label{intro}

Kaon decays, especially some rare kaon decays with ultra small branching ratios, play
an important role in current high-precision test of Standard Model, and provide
excellent channels to probe physics beyond the Standard
Model~\cite{Cirigliano:2011ny}. 
The experimental and theoretical studies of kaon
decays are believed to be more and more important nowadays because kaon decays
have both theoretically clean branching ratios in experimental searches and
gradually improved Standard Model predictions~\cite{Buras:2018wmb}. 

As a typical rare decay, $K\to \ell\nu_\ell \ell'^+ \ell'^-$ involves the
second-order
electroweak interaction, providing a good place to test Standard
Model predictions. In experiments, three types of $K\to \ell\nu_\ell \ell'^+
\ell'^-$ decays have been observed: $K\to e\nu_e e^+ e^-$, $K\to \mu\nu_\mu e^+
e^-$ and $K\to e\nu_e \mu^+ \mu^-$, with small branching ratios in the order of
$O(10^{-8})$~\cite{Poblaguev:2002ug,Ma:2005iv}. 

In theoretical study, the determination of $K\to \ell\nu_\ell \ell'^+ \ell'^-$
decay width is highly nontrivial due to the non-perturbative nature of kaon
internal structure. Since the phase space of $K\to \ell\nu_\ell \ell'^+ \ell'^-$
allows the virtual photon carry relatively large momentum, e.g. the momentum close to the
kaon mass, understanding the momentum
dependence of the decay amplitude is essential in theoretical calculation of
$K\to \ell\nu_\ell \ell'^+ \ell'^-$ decay width. To be specific, apart from the decay constant which describes the point-like interaction, four form factors are involved to
describe the structure-dependent contribution in $K\to
\ell\nu_\ell \ell'^+ \ell'^-$~\cite{Bijnens:1992en}. 
The momentum dependence of these form factors is non-negligible.
In experiments it produced different results by treating the form factors as constants or 
considering the relevant momentum dependence through a simple
vector-meson-dominance model~\cite{Poblaguev:2002ug}. 
Therefore, properly including the momentum dependence of the decay amplitude is important for 
both theoretical predictions and experimental measurements. 

Theoretical study of $K\to \ell\nu_\ell \ell'^+ \ell'^-$ amplitude and form
factors has been carried out using chiral perturbation theory
(ChPT)~\cite{Bijnens:1992en}, where form factors are estimated at the next to leading
order (NLO), and predictions close to experimental results are obtained. 
However, one should note that the form factors $F_V$ and $F_A$ are treated as constants 
at NLO~\cite{Bijnens:1992en}, and the momentum dependence only appears at the NNLO~\cite{Geng:2003mt,Ametller:1993hg}. 
Thus, the theoretical uncertainties due to momentum dependence have not been
estimated thoroughly by the past ChPT studies, which leads to a difficulty to directly compare the results
between experiments and ChPT.

As a generic non-perturbative approach, lattice QCD
can help to improve the SM
predictions of $K\to \ell\nu_\ell \ell'^+ \ell'^-$ decay width. In past few years, some rare kaon decays have
been studied successfully using lattice QCD, such as $K^+\to
\pi^+\nu\bar{\nu}$~\cite{Christ:2016eae,Bai:2017fkh,Bai:2018hqu,Christ:2019dxu,Christ:2020hwe} 
and $K\to\pi \ell^+\ell^-$~\cite{Christ:2015aha,Christ:2016mmq}. 
Besides, other processes involving both weak and electromagnetic interactions,
e.g. the radiative corrections to the leptonic and semileptonic decays, have also been
investigated
recently~\cite{Carrasco:2015xwa,Lubicz:2016xro,Giusti:2017dwk,Feng:2020zdc,Seng:2020wjq,Feng:2020mmb,Desiderio:2020oej,Frezzotti:2020bfa,Seng:2020jtz,Ma:2021}. 
It is interesting to have the lattice QCD study extend its horizon to include the $K\to \ell\nu_\ell \ell'^+ \ell'^-$
decay, where the final state involves four daughter particles.

Here we find that a direct lattice QCD calculation of $K\to \ell\nu_\ell \ell'^+ \ell'^-$ decay width
is encountered with the following technical problems.
\begin{enumerate}
	\item General finite-volume effects: In order to calculate the decay width, one
        needs to know the arbitrary momentum dependence of the 
        decay amplitude.
        However, through discrete Fourier transformation the lattice data from a finite-volume box
        can only access discrete momenta. 
        This problem appears as finite-volume effects in the calculation 
        of decay width.
	
	\item Temporal truncation effects: As it is shown in section \ref{section3}, using
        the hadronic function in coordinate space, we perform an
        integral in Euclidean time to obtain the hadronic function with an
        assigned momentum. 
        In the process of $K\to K^*\ell'^+\ell'^-\to
        \ell\nu_\ell\ell'^+\ell'^-$, the $K^*$ in the intermediate state
        carries nonzero momentum and thus the energy of 
        intermediate state is larger than that of
        the initial/final state. As a result, the time integral 
        converges when the integral range approaches to infinity. 
        However, in the soft-photon region where the
        four momentum of the electromagnetic current $(E,\vec{P})$ is close to
        zero, the integral converges very slowly. Since 
        the lattice temporal extent $T$ is finite, we find that the temporal truncation effects are not
        negligible. An extrapolation to infinitely large time extent 
        is required to achieve a precise calculation.
	
	\item Complex calculation procedures: The calculation of $K\to \ell\nu_\ell
        \ell'^+ \ell'^-$ decay width is of
        particular complication because it involves several form factors and
        four-body phase-space integral. One needs to construct a reliable and
        convenient approach to calculate the decay amplitude at arbitrary
        momenta and perform the
        phase-space integral.
    \item Specific power-law finite-volume effects associated with
        $K\to\pi\pi\ell\nu_\ell\to\ell\nu_\ell\ell'^+\ell'^-$: This subprocess is
        essentially a long-distance process involving multihadron in the
        intermediate state. When the momentum of the electromagnetic current is fixed, 
        the corresponding power-law finite-volume effects have been
        studied first by Ref.~\cite{Christ:2015pwa} using 
        the $K_L$-$K_S$ mass difference as an example 
        and later by Ref.~\cite{Briceno:2019opb} for more general cases.
        When calculating the decay width, the momentum of the
        electromagnetic current runs over the whole allowed phase-space region, the
        finite-volume correction becomes more complicated and still remains an
        open problem. Thus situation also happens for the $K\to \mu^+\mu^-$ decay
        where two off-shell photons are involved~\cite{Christ:2020bzb}.
\end{enumerate}

This work is aiming at solving the first three technical problems, building a
convenient calculation procedure and presenting the lattice results
of $K\to \ell\nu_\ell \ell'^+ \ell'^-$ decay width. The central part
of this paper is to introduce the following methodologies:
1) a scalar function method to compute the hadronic function,
2) infinite volume reconstruction (IVR) method\cite{Feng:2018qpx} to reduce the unphysical 
temporal truncation and finite-volume effects, and 3)
convenient phase-space integration method to obtain the decay width. 

\begin{table*}
	\begin{ruledtabular}
	\begin{tabular}{ccccc}
		Channels& $m_{ee}$ cuts & Lattice ($m_\pi=352$ MeV) &ChPT\cite{Bijnens:1992en} & experiments\\
		\hline
        $\operatorname{Br}[K\to e\nu_e e^+ e^-]$&$140$ MeV&$1.77(16)\times 10^{-8}$&$3.39\times 10^{-8}$&$2.91(23)\times 10^{-8}$\cite{Poblaguev:2002ug}\\
        $\operatorname{Br}[K\to \mu\nu_\mu e^+ e^-]$&$140$ MeV&$10.59(33)\times 10^{-8}$&$8.51\times 10^{-8}$&$7.93(33)\times 10^{-8}$\cite{Poblaguev:2002ug}\\
        $\operatorname{Br}[K\to e\nu_e \mu^+ \mu^-]$&——&$0.72(5)\times 10^{-8}$&$1.12\times 10^{-8}$&$1.72(45)\times 10^{-8}$\cite{Ma:2005iv}\\
        $\operatorname{Br}[K\to \mu\nu_\mu \mu^+ \mu^-]$&——&$1.45(6)\times 10^{-8}$&$1.35\times 10^{-8}$&——\\
	\end{tabular}
\end{ruledtabular}
    \caption{Comparison of branching ratios of
    $\operatorname{Br}[K\to\ell\nu_\ell \ell'^+\ell'^-]$ 
    among our lattice-QCD calculation (at $m_\pi=352$ MeV), ChPT and experiments.
    In order to compare results with ChPT, we choose the same cuts $m_{ee}>140$ MeV 
    as that in Ref.~\cite{Bijnens:1992en}, where $m_{ee}$ is the invariant mass
    of the $e^+e^-$ pair.
    For decays with $e e^+ e^-$, the cuts are applied to both invariant masses.
    (The kaon mass $m_K$ used in the lattice
    calculation is slightly different from the physical kaon mass
    $m_{K,\mathrm{phy}}$. For the lattice results, we rescale the cuts for $m_{ee}$ as
    $\frac{m_{ee}}{m_K}>\frac{140\,\mathrm{MeV}}{m_{K,\mathrm{phy}}}$.) 
    The experimental results of $K\to e\nu_e e^+e^-$ and $K\to \mu\nu_\mu
    e^+e^-$ are the extrapolated values 
    from $m_{ee}>150$ MeV and 145 MeV to $m_{ee}>140$ MeV. The extrapolation
    formula are given in Ref.~\cite{Poblaguev:2002ug}.
    \label{resintro}}
\end{table*}

With these developed methods, we calculate $K\to \ell\nu_\ell \ell'^+
\ell'^-$ decay width using a gauge ensemble of $N_f=2+1+1$-flavor twisted mass fermion
at the unphysical pion mass
$m_\pi=0.3515(15)$ GeV. The valance strange quark mass is tuned to make the kaon
mass $m_K=0.5057(13)$ GeV close to its physical value. 
The lattice results 
of the branching ratios are summarized in Table~\ref{resintro} and are found to
be comparable to experimental measurements and ChPT predictions.
Systematic errors of our lattice calculation mainly come from unphysical quark mass, non-zero lattice spacing and residual finite volume effects. Calculation at the physical quark mass together with the continuum extrapolation will be included in our future work.  
\footnote{When this paper is under peer review, a parallel lattice
study is performed to compute the decay width based on the extraction of the form factors~\cite{Gagliardi:2022szw}.}

In this paper we will first introduce in Sec.~\ref{section2} the decay amplitude of $K\to
\ell\nu_\ell \ell'^+\ell'^-$ in Minkowski space. This part follows
Refs.~\cite{Bijnens:1992en,Kampf:2018wau} and is also a necessary part of lattice calculation. 
In Sec.~\ref{section3} we will establish a connection between Minkowski hadronic function and
Euclidean one, and give more detailed description of the computational techniques mentioned above, 
which is the most central part of this paper. Finally, we present the numerical
results in Sec.~\ref{section4} and reach a conclusion in Sec.~\ref{section5}.

\section{\label{section2}Decay width of $K\to \ell\nu_\ell \ell'^+ \ell'^-$}
Our program aims at the calculation of the branching ratios of $K\to
\ell\nu_\ell \ell'^+ \ell'^-$ via
\begin{equation}\label{Br}
    \operatorname{Br}\left[K\to \ell\nu_\ell \ell'^+ \ell'^-\right]=\frac{1}{2
    m_K \Gamma_{K}} \int d \Phi_{4}\left|\mathcal{M}(K\to \ell\nu_\ell \ell'^+
    \ell'^-)\right|^{2},
\end{equation}
where $\Gamma_K=5.3168(86)\times10^{-14}$ MeV is the kaon decay width from
the Particle Data Group~\cite{Zyla:2020zbs}.\footnote{ Note that we do not calculate the total kaon decay width from lattice.}
$\mathcal{M}(K\to \ell\nu_\ell \ell'^+\ell'^-)$ is the decay amplitude and $\int
d\Phi_4$ indicates a four-body phase-space integral.

Our approach to calculate the $K\to \ell\nu_\ell \ell'^+ \ell'^-$ decay width
includes three major steps. The first step is to determine the Minkowski hadronic functions
        \ba
        &&H_M^{\nu}(q)=\langle 0|J_{W,M}^{\nu}(0)|K(q)\rangle,
        \nonumber\\
        &&H_M^{\mu\nu}(p,q)=\int d^4x\,e^{ip\cdot x}\left\langle
        0\left|T\left\{J_{\text{em},M}^{\mu}(x)
        J_{W,M}^{\nu}(0)\right\}\right| K(q)\right\rangle,
        \ea
where the electromagnetic and weak currents in Minkowski space are defined as
        $J_{\text{em},M}^{\mu}=\frac{2}{3} \bar{u} \gamma^{\mu} u-\frac{1}{3} \bar{d}
        \gamma^{\mu} d-\frac{1}{3} \bar{s} \gamma^{\mu} s$ and $J_{W,M}^{\nu}=\bar{s} \gamma^{\nu}\left(1-\gamma_{5}\right) u$. 
        $p=(E,\vec{p})$ and $q=(m_K,\vec{0})$ are Minkowski 4-momentum of the
        electromagnetic current and initial kaon state. We define the parameters
        $\rho_1$ and $\rho_2$ as
\begin{equation}
\label{pqM}
p^2=\rho_1 m_K^2, \quad (q-p)^2=\rho_2 m_K^2.
\end{equation}
        In a lattice QCD study, the hadronic functions are generally calculated in
        Euclidean space. The connection between Minkowski and Euclidean hadronic
        function is established in Sec.~\ref{section3}.

As a second step, the decay amplitude $\mathcal{M}(K\to \ell\nu_\ell
        \ell'^+ \ell'^-)$  is
        constructed by combining the hadronic function $H_M^{\mu\nu}(p,q)$ with
        the leptonic factor~\cite{Bijnens:1992en}. Here we target on the
        determination of the amplitude $\mathcal{M}$ with arbitrary momentum
        dependence.

As a last step, the decay amplitude is used as an input in the integral~(\ref{Br}) 
        to obtain the decay width. The definition of four-body
        phase-space integral is provided in Ref.~\cite{Kampf:2018wau}, which
        is originally used for the process of $K\to \ell\bar{\ell} \ell'\bar{\ell}'$.
        We use the Monte Carlo method to perform the phase-space integration.

\subsection{Hadronic function in Minkowski space}
In the continuum theory the hadronic function $H_M^{\mu\nu}(p,q)$ satisfies Ward
identity~\cite{Bijnens:1992en}
\begin{equation}
\label{Ward}
p_\mu H_M^{\mu\nu}(p,q)=f_K q^\nu
\end{equation}
with $f_K$ the kaon decay constant. Using Ward identity, 
$H_M^{\mu\nu}(p,q)$ can be written in terms of form
factors~\cite{Desiderio:2020oej} as
\begin{eqnarray}
\label{Hfac}
H_M^{\mu\nu}(p,q)=&&H_{1}\left[p^{2} g^{\mu \nu}-p^{\mu} p^{\nu}\right]+H_{2}\left[\left(p \cdot q-p^{2}\right) p^{\mu}-p^{2}(q-p)^{\mu}\right](q-p)^{\nu}\nonumber\\
&&-i \frac{F_{V}}{m_K} \varepsilon^{\mu \nu \alpha \beta} p_{\alpha} q_{\beta}+\frac{F_{A}}{m_{K}}\left[\left(q \cdot p-p^{2}\right) g^{\mu \nu}-(q-p)^{\mu} p^{\nu}\right]\nonumber \\
&&+f_{K}\left[g^{\mu \nu}+\frac{(2 q-p)^{\mu}(q-p)^{\nu}}{2 q \cdot
    p-p^{2}}\right].
\end{eqnarray}

\begin{figure}[htbp]
	\centering
	\includegraphics[width=0.5\textwidth]{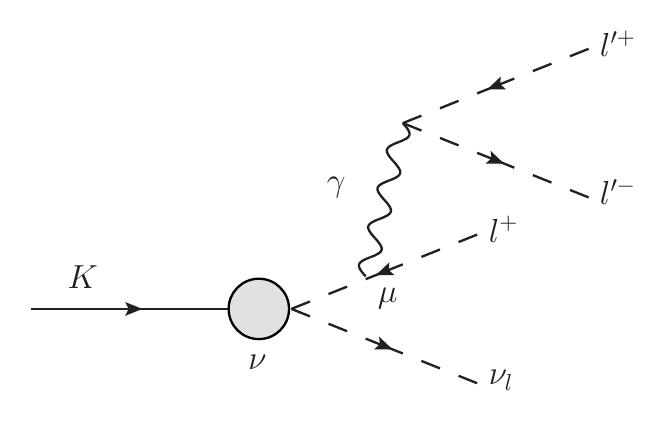}
	\caption{Contribution of off-shell photon radiation from the final-state
    lepton in $K\to \ell\nu_\ell \ell'^+ \ell'^-$. The hadronic part is
    described by $H_M^\nu(q)$.\label{fig:IB}}
\end{figure}

\begin{figure}[htbp]
	\centering
	\includegraphics[width=0.75\textwidth]{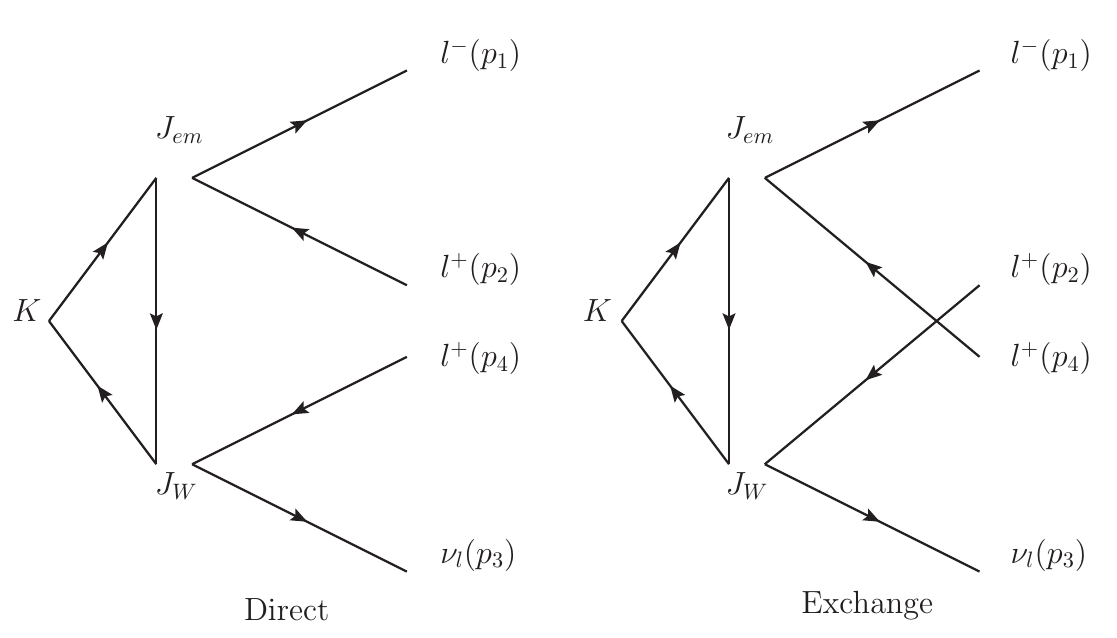}
	\caption{Contribution of off-shell photon radiation from quarks in $K\to
    \ell\nu_\ell \ell^+ \ell^-$.
    The hadronic part is described by $H_M^{\mu\nu}(p,q)$.\label{fig:Fll}}
\end{figure}

Using the hadronic function $H_M^\nu(q)$, one can construct the amplitude
for the subprocess of $K\to \ell\nu_\ell\to \ell\nu_\ell\ell'^+\ell'^-$ as shown in
Fig.~\ref{fig:IB}.
Using the hadronic function $H_M^{\mu\nu}(p,q)$, the remaining contribution to
the decay amplitude
of $K\to \ell\nu_\ell \ell'^+ \ell'^-$ can be constructed. 
For the case of $\ell'=\ell$ the decay amplitude
consists of two parts, shown as the ``Direct'' and ``Exchange'' diagrams in
Fig.~\ref{fig:Fll}.
For $\ell\neq \ell'$, only ``Direct'' diagram contributes.

Here we use the case of $\ell'=\ell$ to introduce the expressions for the decay
amplitudes.
The 4-momenta of the final-state leptons are defined as $p_i$ with $i=1,2,3,4$ as shown
in Fig.~\ref{fig:Fll}. The decay amplitudes are given as
\ba
\label{M}
i \mathcal{M}_{D} &=&-i \frac{G_{F} e^{2} V_{u s}^{*}}{\sqrt{2}
s_{12}}\left[f_{K} L^{\mu}\left(p_{1}, p_{2}, p_{3}, p_{4}\right)-H_M^{\mu
\nu}\left(p_{12}, q\right) l_\nu\left(p_{3},
p_{4}\right)\right]\left[\bar{u}\left(p_{1}\right) \gamma_{\mu}
v\left(p_{2}\right)\right],
\nonumber
\\
i \mathcal{M}_{E} &=&+i \frac{G_{F} e^{2} V_{u s}^{*}}{\sqrt{2}
s_{14}}\left[f_{K} L^{\mu}\left(p_{1}, p_{4}, p_{3}, p_{2}\right)-H_M^{\mu
\nu}\left(p_{14}, q\right) l_\nu\left(p_{3},
p_{2}\right)\right]\left[\bar{u}\left(p_{1}\right) \gamma_{\mu}
v\left(p_{4}\right)\right],
\ea
where the terms with a factor of $f_K$ arise from Fig.~\ref{fig:IB}, which has both the ``Direct'' and the ``Exchange'' contribution if $l = l'$, and the
terms with a
factor of $H_M^{\mu\nu}$ come from Fig.~\ref{fig:Fll}. Here 
$\mathcal{M}_{D}$ and $\mathcal{M}_{E}$ stand for the amplitudes from the
``Direct'' and ``Exchange'' diagrams, respectively.
The photon momentum is given by $p_{ij}\equiv p_i+p_j$ and $s_{ij}\equiv p_{ij}^2$ is 
the momentum square.
The leptonic factors $L^\mu$ and $l^\mu$ are defined as
\begin{eqnarray}
    \label{eq:lepton1}
    &&l^{\mu}\left(p_{3}, p_{4}\right)=\bar{u}\left(p_{3}\right)
    \gamma^{\mu}\left(1-\gamma_{5}\right) v\left(p_{4}\right),
    \nonumber\\
    &&L^{\mu}\left(p_{1}, p_{2}, p_{3},
    p_{4}\right)=l^{\mu}\left(p_{3},p_{4}\right)+{L'}^{\mu}(p_1,p_2,p_3,p_4),
\end{eqnarray}
with
\begin{equation}
    \label{eq:lepton2}
{L'}^{\mu}\left(p_{1}, p_{2}, p_{3}, p_{4}\right)=m_{\ell} \bar{u}\left(p_{3}\right)\left(1+\gamma_{5}\right)
    \frac{2 p_{4}^{\mu}+\slashed {p}_{12}
    \gamma^{\mu}}{m_{\ell}^{2}-\left(p_{4}+p_{12}\right)^{2}}
    v\left(p_{4}\right).
\end{equation}
Note that $\bar{u}$ and $v$ in Eqs.~(\ref{eq:lepton1}) and (\ref{eq:lepton2}) stand for the spinors of $\ell$ and
$\nu_\ell$, which form a charged weak current,
while $\bar{u}$ and $v$ in Eq.~(\ref{M}) stand for the spinors of $\ell^+$ and
$\ell^-$ from an electromagnetic current.
Finally, $V_{us}^*$ is the CKM matrix element and $G_F$ is the Fermi constant. 

It should be noticed that, the term $f_K g^{\mu\nu}$ in Eq.~(\ref{Hfac}) 
can produce a contribution proportional to $f_Kl^\mu$, which exactly cancels 
the $f_Kl^\mu$ term contained by $f_KL^\mu$ in Eq.~(\ref{M}).
We find that the $f_Kl^\mu$ term from Fig.~\ref{fig:IB} would be IR divergent 
in the limit of vanishing lepton mass. 
In order to maintain the exact cancellation in the large IR contribution and
reduce the statistical uncertainty,
we replace $H_M^{\mu\nu}$ by $H_M'^{\mu\nu}$ in Eq.~\ref{M}, with
$H_M'^{\mu\nu}$ defined as
\begin{equation}
    H_M'^{\mu\nu}(p,q)\equiv H_M^{\mu\nu}(p,q)-f_K
    g^{\mu\nu}=H_M^{\mu\nu}(p,q)-\frac{p_\rho H_M^{\rho 0}(p,q)}{m_K}
    g^{\mu\nu}.
\end{equation}
In this way, the amplitude $\mathcal{M}_{D}$ and $\mathcal{M}_{E}$ can be
written as
\begin{eqnarray}
\label{Msub}
    i \mathcal{M}_{D} &=&-i \frac{G_{F} e^{2} V_{u s}^{*}}{\sqrt{2}
    s_{12}}\left[f_{K} L'^{\mu}\left(p_{1}, p_{2}, p_{3}, p_{4}\right)-H_M'^{\mu
    \nu}\left(p_{12}, q\right) l_{\nu}\left(p_{3},
    p_{4}\right)\right]\left[\bar{u}\left(p_{1}\right) \gamma_{\mu}
    v\left(p_{2}\right)\right], 
\nonumber\\
    i \mathcal{M}_{E} &=&+i \frac{G_{F} e^{2} V_{u s}^{*}}{\sqrt{2}
    s_{14}}\left[f_{K} L'^{\mu}\left(p_{1}, p_{4}, p_{3}, p_{2}\right)-H_M'^{\mu
    \nu}\left(p_{14}, q\right) l_{\nu}\left(p_{3},
    p_{2}\right)\right]\left[\bar{u}\left(p_{1}\right) \gamma_{\mu}
    v\left(p_{4}\right)\right].
\end{eqnarray}

Using $\mathcal{M}_{D}$ and $\mathcal{M}_{E}$ as input, the branching ratio of $K\to \ell\nu_\ell
\ell'^+ \ell'^-$ for $\ell=\ell'$ can be calculated through
\begin{equation}
\label{dacay_width}
    \operatorname{Br}[K\to \ell\nu_\ell \ell^+ \ell^-]=\frac{1}{2 m_K \Gamma_{K}} \int d
    \Phi_{4}\left(\left|\mathcal{M}_D\right|^{2}+\left|\mathcal{M}_E\right|^{2}+2\operatorname{Re}[\mathcal{M}_D\mathcal{M}^*_E]\right).
\end{equation}
For $\ell\neq \ell'$, we only have the $\left|\mathcal{M}_D\right|^{2}$ term in the above equation:

\begin{equation}
\label{dacay_width2}
    \operatorname{Br}[K\to \ell\nu_\ell \ell'^+ \ell'^-]=\frac{1}{2 m_K \Gamma_{K}} \int d
    \Phi_{4} \left|\mathcal{M}_D\right|^{2}.
\end{equation}

\subsection{Phase-space integral}
The definition of four-body phase-space integral follows Ref.~\cite{Kampf:2018wau}. In Ref.~\cite{Kampf:2018wau} the formulae are simplified for the case of the daughter
particles with the same masses. Here we generalize the formulae to the case that the daughter
particles have different masses. 

The four-body phase-space $d\Phi_{4}$ is defined as:
\begin{equation}
d \Phi_{4}=\frac{\mathcal{S} \lambda m_K^{4}}{2^{14} \pi^{6}} d x_{12} d x_{34} d
    y_{12} d y_{34} d \phi.
\end{equation}
Here $\mathcal{S}$
is a symmetry factor with $\mathcal{S}=1$ for the case of $\ell\neq\ell'$ and
$\mathcal{S}=\frac{1}{2}$ for  $\ell=\ell'$.
The phase-space variables $x_{12}$ 
$x_{34}$, $y_{12}$, $y_{34}$ and $\phi$ are five independent Lorentz invariant
quantities with $x_{ij}$ and $y_{ij}$ defined as
\begin{equation}
    x_{12}=\frac{s_{12}}{m_K^2},\quad x_{34}=\frac{s_{34}}{m_K^2},\quad
    y_{12}=\frac{2\bar{p}_{12}\cdot p_{34}-2p_{12}\cdot p_{34}\delta_{12}}{\lambda m_K^2},\quad
    y_{34}=\frac{2\bar{p}_{34}\cdot p_{12}-2p_{12}\cdot p_{34}\delta_{34}}{\lambda m_K^2},
\end{equation}
where $\bar{p}_{ij}\equiv p_i-p_j$,
$\lambda\equiv\sqrt{(1-x_{12}-x_{34})^2-4x_{12}x_{34}}$ and $\delta_{ij}\equiv \frac{m_i^2-m_j^2}{s_{ij}}$.
The indices
$1,2,3,4$ specify the particles in the final state. The quantity $\phi$
can be expressed as
\begin{equation}
    \varepsilon_{\mu\nu\rho\sigma}p_1^\mu p_2^\nu p_3^\rho p_4^\sigma
    =-\frac{\lambda
    m_K^4\omega}{16}\sqrt{(\lambda_{12}^2-y_{12}^2)(\lambda_{34}^2-y_{34}^2)}\sin\phi
\end{equation}
with
\begin{equation}
\lambda_{ij}=\sqrt{\left(1-\frac{m_i^2}{s_{ij}}-\frac{m_j^2}{s_{ij}}\right)^2-4\frac{m_i^2m_j^2}{s_{ij}^2}},
    \quad \omega=2\sqrt{x_{12}x_{34}}.
\end{equation}
To create a Monte Carlo generator, it is useful to assign each particle a
4-momentum in the rest frame of kaon in terms of the phase-space variables
as
\ba
\label{eq:momentum_setup}
&&E_{1(2)}=m_K\frac{(1+\delta)(1\pm\delta_{12})\pm \lambda y_{12}}{4},\quad
E_{3(4)}=m_K\frac{(1-\delta)(1\pm\delta_{34})\pm \lambda y_{34}}{4},
\nonumber\\
&&\vec{p}_{1(2)}=\mp
m_K\sqrt{\frac{x_{12}}{4}(\lambda_{12}^2-y_{12}^2)}\hat{x}+m_K\frac{\lambda(1\pm\delta_{12})\pm(1+\delta)y_{12}}{4}\hat{y},
\nonumber\\
&&\vec{p}_{3(4)}=m_K\sqrt{\frac{x_{34}}{4}(\lambda_{34}^2-y_{34}^2)}(\mp\cos\phi\hat{x}\pm\sin\phi\hat{z})
-m_K\frac{\lambda(1\pm\delta_{34})\pm(1-\delta)y_{34}}{4}\hat{y},
\ea
with
$\delta\equiv x_{12}-x_{34}$.
The range of the
phase-space variables are adjusted as
\ba
&&\left(\frac{m_1+m_2}{m_K}\right)^2 \leq x_{12} \leq
\left(1-\frac{m_3+m_4}{m_K}\right)^2,
\nonumber\\
&&\left(\frac{m_3+m_4}{m_K}\right)^2 \leq x_{34} \leq \left(1-\sqrt{x_{12}}\right)^2, 
\nonumber\\
&&-\lambda_{i j} \leq y_{i j} \leq \lambda_{i j}, \quad 0 \leq \phi \leq 2 \pi.
\ea

\section{\label{section3}Methodologies of lattice calculation}
The $K\to\ell\nu_\ell\ell'^+\ell'^-$ decay involves several form factors as given
in Eq.~\ref{Hfac}. The classification of these form factors requires the
constraint from 
Ward identity. In the lattice calculation, Ward identity can be easily violated 
either by the lattice artifacts, e.g. due to the usage of local vector current, or
by the finite-volume effects, e.g. due to the usage of the arbitrary momentum.
These systematic effects significantly affect the precise determination of the form
factors from lattice QCD.
Note that our target is to calculate the total decay width and the
determination of each individual form factor is not a necessary step.
In this work we develop an approach called {\em scalar function method}, which
provides a convenient way to represent the lattice results for coordinate space matrix elements.
Momentum space matrix elements can be obtained from the scalar function
representation with automatic rotational averaging.
The IVR method is then applied to
make the corrections of the temporal truncation effects and finite-volume
effects for the decay amplitude. More details of the methodologies are given as follows.

\subsection{\label{section3.1}Construction of Minkowski hadronic function using
lattice data}
In order to reproduce Minkowski hadronic function using Euclidean lattice data, we shall first 
establish the relation between the hadronic functions in Euclidean and Minkowski
spacetime.

In Euclidean spacetime, the hadronic function is defined as:
\begin{equation}
H_E^{\mu\nu}(x) =\left\langle 0\left|T\left\{J_\text{em}^{\mu}(x)
    J_{W}^{\nu}(0)\right\}\right| K(Q)\right\rangle, 
\end{equation}
where $Q=(im_K,\vec{0})$ is the Euclidean 4-momentum of initial kaon state.
$H_E^{\mu\nu}(x)$ can be extracted from a three-point correlation function $C^{\mu\nu}(\vec{x},t;\Delta T)$ 
\begin{equation}
\label{C3}
C^{\mu \nu}\left(\vec{x}, t ; \Delta T\right)=
    \begin{cases}
\left\langle J_\text{em}^{\mu}(\vec{x}, t) J_{W}^{\nu}(\vec{0}, 0)
        \phi_{K}^{\dagger}(-\Delta T)\right\rangle,  & t \geq 0, \\
\left\langle J_{W}^{\mu}(\vec{0}, 0) J_\text{em}^{\nu}(\vec{x}, t)
        \phi_{K}^{\dagger}(t-\Delta T)\right\rangle, & t<0.
    \end{cases}
\end{equation} 
We choose sufficiently large $\Delta T$ to guarantee kaon ground-state dominance. 
Then the hadronic function
$H^{\mu\nu}_E(\vec{x},t)$ can be determined through
\begin{equation}
\label{H}
H_{E,A / V}^{(L),\mu \nu}(\vec{x}, t)=
    \begin{cases}
        N_{K}^{-1} Z_{V} Z_{A / V} e^{m_{K} \Delta T} C_{A / V}^{\mu
        \nu}(\vec{x}, t,\Delta T), & t\geq 0, \\
N_{K}^{-1} Z_{V} Z_{A / V} e^{m_{K}(\Delta T-t)} C_{A / V}^{\mu \nu}(\vec{x},
        t,\Delta T), & t<0.
    \end{cases}
\end{equation}
In Eq.~(\ref{H}) we have separated the weak current into axial-vector-current and vector-current
parts by using the
subscript $A/V$. $Z_A$ and $Z_V$ are the corresponding renormalization factors.
We use the superscript $(L)$ to emphasize that this hadronic function is calculated
in the finite volume. The normalization factor
$N_{K}=\langle K|\phi_{K}^{\dagger}(0)| 0\rangle/(2 m_{K})$ and the kaon mass
$m_K$ can be
calculated from the lattice two-point functions. 

For simplicity, let us first consider the infinite-volume Euclidean hadronic function
$H_E^{\mu\nu}(x)$. 
In momentum space it is given by
\begin{equation}\label{Hcalc}
    H_{E}^{\mu\nu}(P,Q)=-i\int_{-T/2}^{T/2} dt\int d^3\vec{x}\, e^{Et-i\vec{p}\cdot\vec{x}}H_E^{\mu\nu}(x)
\end{equation}
with the Euclidean momenta
\begin{equation}
    \label{eq:rho1rho2}
    P=(iE,\vec{p}), \quad -P^2=\rho_1 m_K^2, \quad -(Q-P)^2=\rho_2 m_K^2.
\end{equation}

\begin{figure}[htbp]
	\centering
	\includegraphics[width=0.6\textwidth]{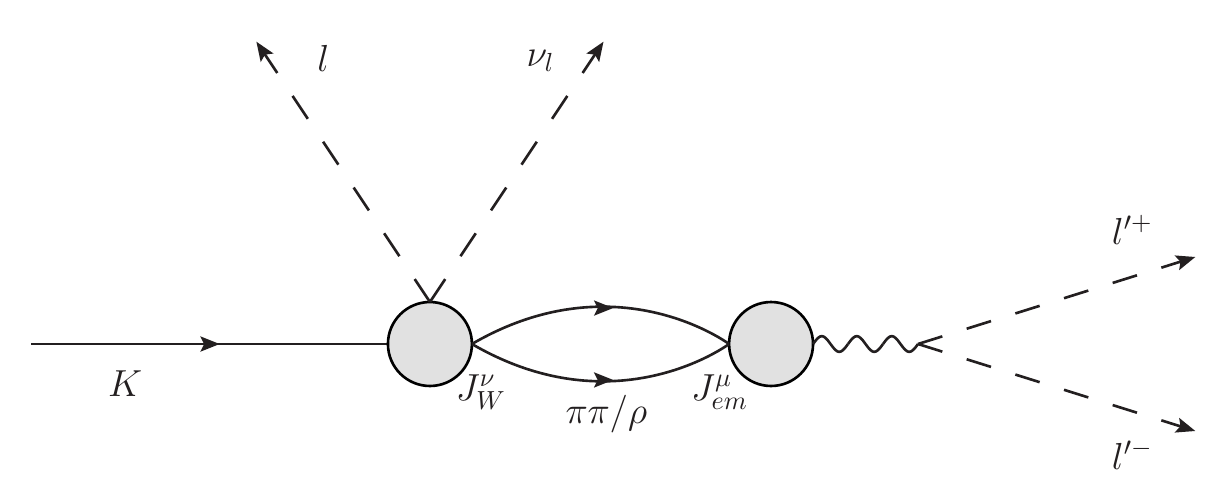}
	\caption{Low-lying-state dominance for $t>0$: $\pi\pi$ or $\rho$ states.\label{fig:pipiint}}
	\includegraphics[width=0.6\textwidth]{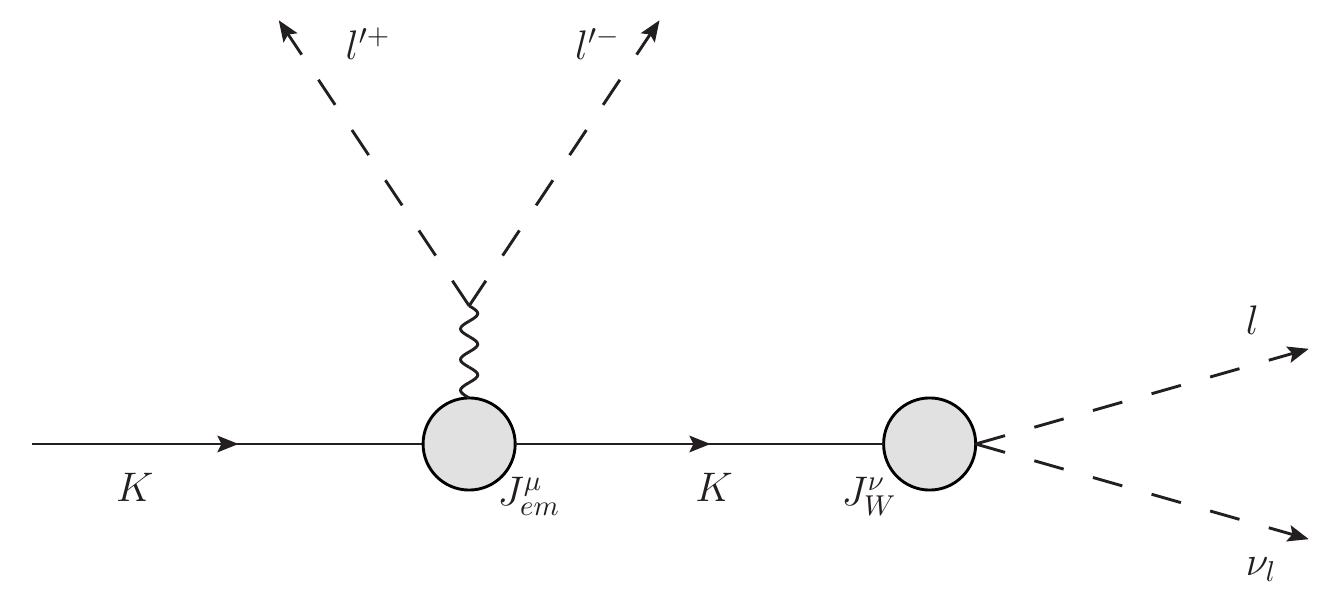}
	\caption{Low-lying-state dominance for $t<0$: kaon states.\label{fig:Kint}}
\end{figure}

In order to calculate decay width, the Euclidean function $H_{E}^{\mu\nu}(P,Q)$
should be related to Minkowski one $H_{M}^{\mu\nu}(p,q)$, which is defined in
Sec.~\ref{section2}. This relation can be established by inserting the complete set
of intermediate states into $H_{M}^{\mu\nu}(p,q)$ through
\begin{eqnarray}\label{Hint1}
    H^{\mu\nu}_M(p,q)&&=\int_0^\infty dt \sum_n\langle
    0|J_{\text{em},M}^\mu(0)|n(\vec{p})\rangle_{M}\langle
    n(\vec{p})|J_{W,M}^\nu(0)|K(q)\rangle_M e^{i(E-E_n+i\epsilon)t}\nonumber\\
    &&+\int_{-\infty}^0 dt \sum_{n_s}\langle
    0|J_{W,M}^\nu(0)|n_s(-\vec{p})\rangle_{M}\langle
    n_s(-\vec{p})|J_{\text{em},M}^\mu(0)|K(q)\rangle_M e^{i(E+E_{n_s}-m_K-i\epsilon)t}\nonumber\\
    &&=i\sum_n\frac{1}{E-E_n+i\epsilon} \langle
    0|J_{\text{em},M}^\mu(0)|n(\vec{p})\rangle_{M}\langle
    n(\vec{p})|J_{W,M}^\nu(0)|K(q)\rangle_M\nonumber\\
&&- i\sum_{n_s} \frac{1}{E+E_{n_s}-m_K-i\epsilon}\langle
    0|J_{W,M}^\nu(0)|n_s(-\vec{p})\rangle_{M}\langle
    n_s(-\vec{p})|J_{\text{em},M}^\mu(0)|K(q)\rangle_M.
\end{eqnarray}
For $t>0$ and $t<0$, the intermediate
states are denoted as the state $|n\rangle$ with strangeness $S=0$ and the state
$|n_s\rangle$ with $S=1$, respectively. 
The matrix elements $\langle \cdots\rangle_{M}$ carry a
subscript $M$, which reminds us that it is defined in Minkowski space. 
The low-lying states for $|n\rangle$ are given by the $p$-wave $\pi\pi$ states,
which couple to the $\rho$ resonance. The lowest state for
$|n_s\rangle$ is given by $|K\rangle$ state. The relevant diagrams for these
low-lying states are shown in
Figs.~\ref{fig:pipiint} and \ref{fig:Kint}. 

When inserting the complete set of intermediate states into the Euclidean
function
$H^{\mu\nu}_E(P,Q)$, we have
\begin{eqnarray}\label{Hint2}
    H^{\mu\nu}_E(P,Q)&&=-i\int_0^{T/2} dt \sum_n\langle
    0|J_{\text{em}}^\mu(0)|n\rangle_{E}\langle n|J_W^\nu(0)|K(Q)\rangle_E e^{(E-E_n)t}\nonumber\\
    &&-i\int_{-T/2}^0 dt \sum_{n_s}\langle
    0|J_{W}^\nu(0)|n_s\rangle_{E}\langle n_s|J_{\text{em}}^\mu(0)|K(Q)\rangle_E e^{(E+E_{n_s}-m_K)t}\nonumber\\
    &&=i\sum_n\frac{1-e^{-(E_n-E)T/2}}{E-E_n} \langle
    0|J_{\text{em}}^\mu(0)|n\rangle_{E}\langle n|J_W^\nu(0)|K(Q)\rangle_E\nonumber\\
&&- i\sum_{n_s} \frac{1-e^{-(E+E_{n_s}-m_K)T/2}}{E+E_{n_s}-m_K}\langle
    0|J_{W}^\nu(0)|n_s\rangle_{E}\langle n_s|J_{\text{em}}^\mu(0)|K(Q)\rangle_E.
\end{eqnarray}
Different from the Minkowski expression, we have introduced a time integral
range $[-T/2, T/2]$ to define the Euclidean hadronic function.
This is because the state $n$
consists of a continuous set of $\pi\pi$ state.
When $E_n=E_{\pi\pi}<E$, the factor
$e^{-(E_n-E)T/2}$ exponentially grows as $T$ increases. In this case, one needs to
use the finite-volume hadronic function $H_E^{(L),\mu\nu}(P,Q)$, where the
low-lying $\pi\pi$ states are discrete. When a spatial momentum
$\vec{p}$ is assigned, one can remove the exponential factor of $e^{-(E_n-E)T/2}$
by isolating each low-lying $\pi\pi$ state. After that,
the difference between
$H_E^{(L),\mu\nu}(P,Q)$ and the real part of $H_M^{\mu\nu}(p,q)$ can be taken into account by 
the
finite-volume correction formula developed in Refs.~\cite{Christ:2015pwa,Briceno:2019opb}. 
The imaginary part of $H_M^{\mu\nu}(p,q)$ can be reproduced by calculating the
on-shell decay amplitudes $K\to \pi\pi \ell \nu$ and $\gamma^*\to \pi\pi$, where
the finite-volume technique is mature~\cite{Lellouch:2000pv,Meyer:2011um}. 
The timelike pion form factor from $\gamma^*\to \pi\pi$
has been calculated on lattice since 2014~\cite{Feng:2014gba,Andersen:2018mau,Erben:2019nmx}.

In this study, we perform the calculation at the unphysical pion mass
$m_\pi=0.3515(15)$ GeV. As a result, $E_{\pi\pi}-E$ is always larger than zero.
In this case one could take the limit of $T\to\infty$ and establish the relation
between Euclidean and Minkowski matrix element. Due to the different convention for the $J^\mu$ operator:
\begin{equation}
\begin{aligned}
&\langle A|J^{\mu}(0)|B\rangle_M\propto (p^\mu_{A}\pm p^\mu_{B}),\quad p^\mu=(p^0,\vec{p}) \\
&\langle A|J^{\mu}(0)|B\rangle_E\propto (P^\mu_{A}\pm P^\mu_{B}),\quad P^\mu=(ip^0,\vec{p})
\end{aligned}
\end{equation}
with $J^\mu$ can be either electromagnetic or weak current, one can verify that
\begin{equation}
    H_E^{\mu\nu}(P,Q)\big|_{T\to\infty}=c^{\mu\nu}H_M^{\mu\nu}(p,q)
\end{equation}
with $c^{00}=-1$, $c^{0i}=c^{i0}=i$ and $c^{ij}=1$.
Thus we have shown that the Minkowski hadronic function can be calculated by
Euclidean lattice data.

\subsection{\label{section3.2}Scalar function method}
In the following part of the paper, for simplicity we will omit the subscript $E$ in the
Euclidean hadronic function and use $H^{\mu\nu}(x)$ and $H^{\mu\nu}(P,Q)$ to replace
$H_E^{\mu\nu}(x)$, $H_E^{\mu\nu}(P,Q)$. It is straightforward to compute
$H^{\mu\nu}(P,Q)$ using the $4\times 4$ Lorentz tensor $H^{\mu\nu}(x)$ as
input
\begin{equation}
\label{Hcalc2}
    H^{\mu\nu}(P,Q)=-i\int_{-T/2}^{T/2} dt\int d^3\vec{x}\,
    e^{Et-i\vec{p}\cdot\vec{x}}H^{\mu\nu}(x).
\end{equation}
We call this method as the {\em direct method}.

In a realistic lattice calculation, $H^{\mu\nu}(x)$ is given by the finite-volume lattice data
$H^{(L),\mu\nu}(x)$. The data depends on $L^3\times T$ spacetime coordinates,
$4\times 4$ Lorentz indices and two types of current insertions ($V/A$). Take a
$24^3\times 48$ lattice as an example, the data size is about 650 MB per
configuration as shown in Table~\ref{scalarmethod}. The computation of the decay width
requires the integration using $H^{(L),\mu\nu}(x)$ as input and is thus quite
complicated.

\begin{table}
	\begin{ruledtabular}
	\begin{tabular}{ccc}
		Method&Direct method&Scalar function method\\
		\hline
		Stored data&$H^{(L),\mu\nu}(x)$& $I^{(L)}_i(|\vec{x}|^2,t)$ \\
		Space-time dimensions&$(L,L,L,T): 24^3\times 48$& $(r^2_{\text{max}},T): 432\times 48$ \\
		Other dimensions&$(\mu,\nu,V/A):4\times 4\times 2$&6\\
		Data size & $\approx 650$ MB/conf & $\approx2$ MB/conf\\
	\end{tabular}
\end{ruledtabular}
	\caption{\label{scalarmethod}Comparison of the size of input data required
    by scalar function method and direct method. By using scalar function
    method, the hadronic function $H^{\mu\nu}(x)$ is converted into six scalar functions $I_i(|\vec{x}|^2,t)$. 
    The former requires the total data size of $L^3\times T\times4\times4\times2$
    while the latter only requires $r_{\mathrm{max}}^2\times T \times 6$ with
    $r_{\mathrm{max}}^2\equiv 3(L/2)^2$. As the lattice size $L$ increases, the
    scalar function method become more efficient compared the direct one.}
\end{table}

In this work we propose to use the scalar function method,
which could significantly reduce the size of data input and provide automatic rotational averaging.
The simplification can be achieved by converting
$H^{\mu\nu}(P,Q)$ and $H^{\mu\nu}(x)$
into the Lorentz scalar functions. For example, $H^{\mu\nu}(P,Q)$ can be used to
construct the following Lorentz invariant quantities
\begin{eqnarray}
    \label{eq:scalar_mom}
    &&\tilde{I}_1(\rho_1,\rho_2)=-\delta^{\mu\nu}m_K^2H^{\mu\nu}(P,Q),\quad
    \tilde{I}_2(\rho_1,\rho_2)=Q^\mu Q^\nu H^{\mu\nu}(P,Q),
    \nonumber\\
    &&\tilde{I}_3(\rho_1,\rho_2)=P^\mu Q^\nu H^{\mu\nu}(P,Q),\quad
    \tilde{I}_4(\rho_1,\rho_2)=Q^\mu P^\nu H^{\mu\nu}(P,Q),
    \nonumber\\
    &&\tilde{I}_5(\rho_1,\rho_2)=P^\mu P^\nu H^{\mu\nu}(P,Q),\quad
    \tilde{I}_6(\rho_1,\rho_2)=\varepsilon^{\mu\nu\alpha \beta}P^\alpha Q^\beta H^{\mu\nu}(P,Q).
\end{eqnarray}
Since the momentum $Q$ satisfies the on-shell condition $Q^2=-m_K^2$, the quantities
$\tilde{I}_i(\rho_1,\rho_2)$ only depend on two variables $\rho_1$ and $\rho_2$, which are
defined in Eq.~(\ref{eq:rho1rho2}). Then we can write $H^{\mu\nu}(P,Q)$ as a
combination of 
\begin{equation}
\label{Scalarmethod1}
    H^{\mu\nu}(P,Q)=\sum_{i=1}^6 \tilde{w}^{\mu\nu}_{i}(P,Q) \tilde{I}_i(\rho_1,\rho_2),
\end{equation}
where $\tilde{w}^{\mu\nu}_{i}(P,Q)$ are analytically known Lorentz factors. The way to obtain $\tilde{w}^{\mu\nu}_{i}(P,Q)$ has been discussed in
Appendix~\ref{B}.

For $H^{\mu\nu}(x)$, we can also write them in terms of Lorentz invariant
quantities through
\begin{equation}
\label{eq:HxQ}
H^{\mu\nu}(x)=\sum_{i=1}^6 w^{\mu\nu}_{i}(x) I_i(|\vec{x}|^2,t),
\end{equation}
where $I_i$ are defined as
\begin{eqnarray}
\label{scalar}
    &&I_{1}(|\vec{x}|^2,t)=\delta^{\mu\nu}H^{\mu\nu}(x),
    \nonumber\\
    &&I_{2}(|\vec{x}|^2,t)=-\frac{Q^\mu Q^\nu}{m_K^2}H^{\mu\nu}(x)=H^{00}(x),
    \nonumber\\
    &&I_{3}(|\vec{x}|^2,t)=\frac{x^\mu Q^\nu}{im_K} H^{\mu\nu}(x)-\frac{x\cdot Q}{im_K} I_2= x^i H^{i0}(x),
\nonumber\\
    &&I_{4}(|\vec{x}|^2,t)=x^i H^{0i}(x),\nonumber\\
    &&I_{5}(|\vec{x}|^2,t)=x^\mu x^\nu  H^{\mu\nu}(x)-\frac{x\cdot
    Q}{im_K}(I_3+I_4)-\left(\frac{x\cdot Q}{im_K}\right)^2 I_2 = x^i x^j
    H^{ij}(x),\nonumber\\
    &&I_{6}(|\vec{x}|^2,t)=\varepsilon^{\mu\nu\alpha 0}x^\alpha H^{\mu\nu}(x).
\end{eqnarray}
It is more convenient to write $I_i$ as the functions of the variables 
$(|\vec{x}|^2,t)=(x^2-(x\cdot Q)^2/(im_K)^2, (x\cdot Q)/(im_K))$.
Again, $w^{\mu\nu}_{i}(x)$ are also the known factors.
The choice of the scalar functions is not unique. Here we design the scalar
functions using the simple combination of $x^\mu$ and $H^{\mu\nu}(x)$.

We then put Eqs.~(\ref{eq:scalar_mom}) and (\ref{scalar}) into
Eq.~(\ref{Hcalc2}) and obtain a relation between
$I_j(|\vec{x}|^2,t)$ and $\tilde{I}_i(\rho_1,\rho_2)$
\begin{equation}
\label{Scalarmethod2}
\tilde{I}_i(\rho_1,\rho_2)=\int d^4 x\, \phi_{ij}(\rho_1,\rho_2;|\vec{x}|^2,t)
    I_j(|\vec{x}|^2,t).
\end{equation}
The detailed expressions for $\phi_{ij}(\rho_1,\rho_2;|\vec{x}|^2,t)$ are given
in Appendix~\ref{B}. Note that in $I_j(|\vec{x}|^2,t)$ the index $j$ belongs to
$1\le j\le6$, $|\vec{x}|^2$ takes values from
$[0,r_{\mathrm{max}}^2\equiv 3(L/2)^2]$ and $t$ ranges from $[-T/2,T/2]$. Thus the total
data size for $I_j(|\vec{x}|^2,t)$ is accounted as
$\sim 6\times r_{\mathrm{max}}^2\times T$, which is significantly smaller than the size of
$H^{\mu\nu}(x)$. A comparison is made in Table~\ref{scalarmethod} to
demonstrate the efficiency of the scalar function method.

Using $I_j(|\vec{x}|^2,t)$ as input and adopting Eqs.~(\ref{Scalarmethod2}),
and (\ref{Scalarmethod1}), the hadronic
function $H^{\mu\nu}(P,Q)$ can be constructed for arbitrary momenta $P$.
 Then the decay amplitude $\mathcal{M}(K\to\ell\nu\ell'^+\ell'^-)$ can be determined. Although $I_j(|\vec{x}|^2,t)$ is calculated within a finite volume and fixed boundary condition, it only causes the exponentially
suppressed finite-volume effects to $\tilde{I}_i(\rho_1,\rho_2)$ due to the cluster decomposition property of QCD.

In the continuum theory
$\tilde{I}_i(\rho_1,\rho_2)$ for $i=1,\cdots6$ are not fully independent
due to the constraint from Ward identity~(\ref{Ward}). As a result,
the scalar functions $\tilde{I}_3(\rho_1,\rho_2)$ and $\tilde{I}_5(\rho_1,\rho_2)$ are both proportional to the decay constant $f_K$ as
\begin{eqnarray}
\label{eq:WI_constraint}
&&\tilde{I}_3(\rho_1,\rho_2)=P^\mu Q^\nu H^{\mu\nu}(P,Q)=m_K^2 f_K
\nonumber\\
&&\tilde{I}_5(\rho_1,\rho_2)=P^\mu P^\nu H^{\mu\nu}(P,Q)=\frac{1+\rho_1-\rho_2}{2}m_K^2 f_K
\end{eqnarray}
One can either use both $\tilde{I}_3(\rho_1,\rho_2)$ and $\tilde{I}_5(\rho_1,\rho_2)$ 
in the calculation or use one of them by treating the other one as the dependent quantity. (In practice, we replace $\tilde{I}_5(\rho_1,\rho_2)$ by $\tilde{I}_3(\rho_1,\rho_2)$.)
In the continuum theory with an infinite volume, the two setups are equivalent. On the lattice,
the results may disagree due to the violation of Ward identity by both lattice artifacts and finite-volume effects. In Sec.~\ref{section4.2}, we calculate the branching ratios using both setups (denoted as ``not using Ward-identity constraint'' and ``using Ward-identity constraint'' respectively) and get consistent results.

\subsection{\label{section3.3}IVR method}

When utilizing Eq.~(\ref{Hcalc2}) to calculate $H^{\mu\nu}(P,Q)$ on lattice, it shall be pointed out that $H^{\mu\nu}(x)$
needs to be replaced by the finite-volume lattice data $H^{(L),\mu\nu}(x)$. As the lattice simulation is performed with a finite temporal extend $T$
and spatial extent $L$, the replacement will cause the temporal truncation effects 
and finite-volume effects. 

In this section, we will use the IVR method~\cite{Feng:2018qpx,Feng:2021zek} to perform the correction for both
temporal truncation and finite-volume effects. 
The idea of the IVR method is that ``infinite-volume data'' $H^{\mu\nu}(x)$ can 
be reconstructed from the finite-volume lattice data $H^{(L),\mu\nu}(x)$.
This method has been successfully applied to
various calculations such as neutrinoless double beta decay~\cite{Tuo:2019bue},
pion charge radius~\cite{Feng:2019geu}, radiative decays~\cite{Meng:2021ecs} and two-photon exchange contributions~\cite{Fu:2022fgh}. 
In this work, the IVR method is separated into two steps, namely IVR and $\delta_{\text{IVR}}$. The former mainly corrects the temporal truncation effects, and the latter focuses on the finite-volume effects.

For simplicity, we will discuss IVR techniques using $H^{\mu\nu}(x)$ and
$H^{\mu\nu}(P,Q)$, and leave IVR
formulae for the scalar functions $I_i(|\vec{x}|^2,t)$ and
$\tilde{I}_i(\rho_1,\rho_2)$ in Appendix~\ref{B}. It should be
reminded that the formulae in Appendix~\ref{B} is what we have actually used in
the numerical calculation.

\subsubsection{Temporal truncation effects}

We start the discussion from the temporal truncation effects. As shown in
Eq.~(\ref{Hint2}), when the temporal extent $T$ increases, the unphysical
terms either exponentially decrease or increase depending on the energy
difference between the intermediate states and initial/final state.
Since these unphysical effects are dominated by the ground-state contributions
as shown in Fig.~\ref{fig:pipiint} and Fig.~\ref{fig:Kint}, here we only
consider the low-lying states $|n\rangle=|\pi\pi(I=L=1)\rangle$ for $t>0$ and $|n_s\rangle=|K\rangle$
for $t<0$. 

For $t>0$, since we use the gauge ensemble with $m_\pi=0.3515(15)$ GeV and
$m_K=0.5057(13)$ GeV, $\pi\pi$ states are always heavier than the kaon state. 
In our numerical study, we do not observe any statistically significant temporal
truncation effects and the unphysical contribution from $e^{-(E_{\pi\pi}-E)T/2}$
can be safely neglected.

For $t<0$, the temporal truncation effects are not negligible especially in the soft photon region, 
where the electromagnetic current carries vanishing momentum $P=(iE,
\vec{p})\approx (0,\vec{0})$. As a result, the intermediate kaon state has the
energy $E_K$ very close to the energy of the initial state $m_K$. A very large
$T$ is required to make the factor $e^{-(E+E_K-m_K)T/2}$ sufficiently small.
Unfortunately, this requirement is not satisfied by a typical lattice temporal
extent of a few fm. Thus the
exponential term $e^{-(E+E_K-m_K)T/2}$ is far from convergence, leading to a large temporal
truncation effects.

In our numerical calculation, we find that even beyond the soft photon region, 
the temporal truncation effects are generally not negligible. 
It means that we need a systematic improved method to reduce the unphysical
contamination from $e^{-(E+E_K-m_K)T/2}$. 

\subsubsection{Finite-volume effects}

As explained in Ref.~\cite{Feng:2018qpx}, the size of the hadronic function $H^{\mu\nu}(x)$ exponentially suppresses at large
spatial separation $|\vec{x}|$. The rate of suppression depends on the energy
difference between intermediate state and the initial state. If
$H^{\mu\nu}(x)$ does not decrease to zero at the boundary of the
lattice, then the finite-volume effects are expected to be non-negligible. 
In other
words, the lattice data $H^{(L),\mu\nu}(x)$ could deviate from the infinite-volume $H^{\mu\nu}(x)$
by a sizable difference. These finite-volume effects propagate into 
$H^{\mu\nu}(P,Q)$ and are more enhanced when $P$ is a non-lattice momentum.

After the explanation of the origin of both temporal truncation and finite-volume
effects, we will start to describe the IVR method to reduce these unphysical systematic effects.

\subsubsection{Step 1: IVR}

In the $t<0$ region, the hadronic function $H^{\mu\nu}(\vec{x},t)$ is saturated
by the single particle states at sufficiently large $|t|$.
We remark such time separation as $|t| > t_s$.
For $t\le -t_s$, the hadronic function can be given by
\begin{eqnarray}
H^{\mu\nu}(\vec{x},t)\big|_{t\leq -t_s}=&&\langle 0| J_A^\nu(\vec{0},0)J_{\text{em}}^\mu(\vec{x},t)|K\rangle\nonumber\\
=&&\int \frac{d^3 p_K}{(2\pi)^32E_K}\langle 0|J_A^\nu(\vec{0},0)|K(p_K)\rangle\langle K(p_K)|J_{\text{em}}^\mu(\vec{0},0)|K\rangle e^{-i\vec{p}_K\cdot \vec{x}}e^{(E_K-m_K)t}\nonumber\\
=&&\int \frac{d^3 p_K}{(2\pi)^3}\tilde{H}^{\mu\nu}(\vec{p}_K,E_K)
    e^{-i\vec{p}_K\cdot \vec{x}}e^{(E_K-m_K)t},
\end{eqnarray}
with $\tilde{H}^{\mu\nu}(\vec{p}_K,E_K)$ defined as
\begin{equation}
\tilde{H}^{\mu\nu}(\vec{p}_K,E_K)=\frac{1}{2E_K}\langle
    0|J_A^\nu(\vec{0},0)|K(p_K)\rangle\langle
    K(p_K)|J_{\text{em}}^\mu(\vec{0},0)|K\rangle.
\end{equation}
We can determine $\tilde{H}^{\mu\nu}(\vec{p}_K,E_K)$ using
$H^{\mu\nu}(\vec{x},t)$ at $t=-t_s$ as an input
\begin{equation}
    \label{eq:ground_state_dominance}
\tilde{H}^{\mu\nu}(\vec{p}_K,E_K)=\int
    d^3x'H^{\mu\nu}(\vec{x}',-t_s)e^{i\vec{p}_K\cdot \vec{x}'}e^{(E_K-m_K)t_s}.
\end{equation}
Using the expression of $\tilde{H}^{\mu\nu}(\vec{p}_K,E_K)$ in Eq.~(\ref{eq:ground_state_dominance}), all the hadronic
function $H^{\mu\nu}(\vec{x},t)$ with $t<-t_s$ can be reconstructed via
\begin{eqnarray}
\label{IVRHx}
H^{\mu\nu}(\vec{x},t)\big|_{t\leq -t_s}
=&&\int \frac{d^3 p_K}{(2\pi)^3}\tilde{H}^{\mu\nu}(\vec{p}_K,E_K)
    e^{-i\vec{p}_K\cdot \vec{x}}e^{(E_K-m_K)t}\nonumber\\=&&\int \frac{d^3
    p_K}{(2\pi)^3}\int d^3x'H^{\mu\nu}(\vec{x}',-t_s)e^{i\vec{p}_K\cdot
    (\vec{x}'-\vec{x})}e^{(E_K-m_K)(t+t_s)}.
\end{eqnarray}

As a next step,
the time integral~(\ref{Hcalc2}) with the range $-T/2<t<0$ can be separated into two parts: $-t_s<t<0$ and
$-T/2<t<-t_s$. We can extend the lower bound of the integral from $-T/2$ to
$-\infty$.
By putting Eq.~(\ref{IVRHx}) into the integral, we have
\begin{eqnarray}
\label{IVRH}
    &&\int^0_{-\infty} dt \int d^3\vec{x}\, e^{Et-i\vec{p}\cdot\vec{x}}H^{\mu\nu}(\vec{x},t)\nonumber\\
    =&&\int_{-t_s}^{0} dt \int d^3\vec{x}\,
    e^{Et-i\vec{p}\cdot\vec{x}}H^{\mu\nu}(\vec{x},t)+\int_{-\infty}^{-t_s} dt
    \int d^3\vec{x}\, e^{Et-i\vec{p}\cdot\vec{x}}H^{\mu\nu}(\vec{x},t)\nonumber\\
    =&&\int_{-t_s}^{0} dt \int d^3\vec{x}\,
    e^{Et-i\vec{p}\cdot\vec{x}}H^{\mu\nu}(\vec{x},t)+\int d^3\vec{x}\,
    e^{-i\vec{p}\cdot\vec{x}}H^{\mu\nu}(\vec{x},-t_s)\frac{e^{-Et_s}}{E+E_K-m_K}.
\end{eqnarray}
From the second to the third line, the hadronic function $H^{\mu\nu}(\vec{x},t)$
with $t<-t_s$ is reconstructed using Eq.~(\ref{IVRHx}). 
Using the hadronic function at some modest value of $t_s$, it
allows us to perform the time integral analytically in the whole region of $-\infty <t<-t_s$.
Thus the temporal truncation effects naturally disappear.

In the practical calculation, we use the IVR method to reconstruct the scalar
functions
$\tilde{I}_i(\rho_1,\rho_2)$.
The treatment is very similar as described above. We separate the time integral
into the short-distance part with $t>-t_s$ and long-distance part with $t<-t_s$ and 
obtain $\tilde{I}_{i}^{(s)}(\rho_1,\rho_2)$ and
$\tilde{I}_{i}^{(l)}(\rho_1,\rho_2)$ correspondingly. The total contribution is
a combination of 
\begin{equation}\label{IVRI}
    \tilde{I}_i^\text{IVR}(\rho_1,\rho_2)=\tilde{I}_{i}^{(s)}(\rho_1,\rho_2)+\tilde{I}_{i}^{(l)}(\rho_1,\rho_2),\quad
    i=1,\cdots,6.
\end{equation}
The detailed expressions for $\tilde{I}_{i}^{(s)}$ and $\tilde{I}_{i}^{(l)}$ are
given in Appendix~\ref{B}.

\subsubsection{Step 2: $\delta_\text{IVR}$}

After making correction to the temporal truncation effects, the lattice results 
are still affected by the finite-volume effects, which is denoted as
$\delta_\text{IVR}$ here. In
Ref.~\cite{Feng:2018qpx}, such correction has been demonstrated to be
exponentially suppressed as the lattice size $L$ increases. This exponential
behavior is also confirmed by our numerical analysis in Sec.~\ref{section4}. 
We can calculate and correct the leading effect of this already exponentially suppressed
finite volume error.
Although exponentially suppressed, $\delta_\text{IVR}$ can still be very large
if a relatively small lattice (e.g. $La=2.2$ fm in this calculation) is used.

We first introduce the definition of the correction term $\delta_\text{IVR}$ as follows. 
The hadronic function $H^{\mu\nu}(P,Q)$ is calculated using $H^{(L),\mu\nu}(x)$ as an input: 
\begin{eqnarray}\label{DH}
H^{\mu\nu}(P,Q)&&=\int d^4 x\, e^{Et-i\vec{p}\cdot\vec{x}}H^{\mu\nu}(x)\nonumber\\
&&=\int_V d^4 x\, e^{Et-i\vec{p}\cdot\vec{x}}H^{(L),\mu\nu}(x)
    \nonumber\\
    &&    +\int_V d^4 x\,
    e^{Et-i\vec{p}\cdot\vec{x}}\left(H^{\mu\nu}(x)-H^{(L),\mu\nu}(x)\right)
    \nonumber\\
    &&+\int_{>V} d^4 x\, e^{Et-i\vec{p}\cdot\vec{x}}H^{\mu\nu}(x).
\end{eqnarray}
In the above equation, $\int_V d^4x$ indicates that the integral is carried out
within a finite spatial volume, while $\int_{>V} d^4x$ means that the integral is
performed outside the lattice box in the spatial directions. The second line of Eq.~(\ref{DH}) shows
the contribution from lattice data $H^{(L),\mu\nu}(x)$. The remaining
contributions are given by the third and fourth lines in Eq.~(\ref{DH}) and denoted as
$\delta_\text{IVR}^{(1)}$ and $\delta_\text{IVR}^{(2)}$, respectively. Combining
$\delta_\text{IVR}^{(1)}$ and $\delta_\text{IVR}^{(2)}$ together yields the so-called correction
$\delta_\text{IVR}$. 

\begin{figure}
	\centering
	\includegraphics[width=0.5\textwidth]{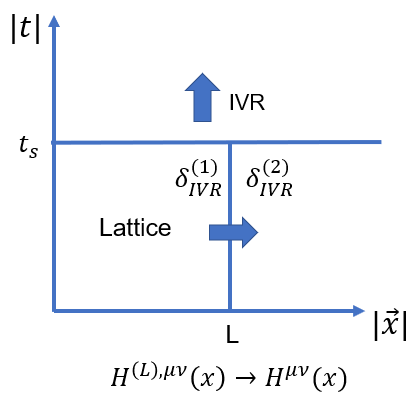}
	\caption{The main idea of IVR method\label{IVR}. In the temporal direction,
    $H^{\mu\nu}(x)\big|_{t<-t_s}$ is reconstructed by using
    $H^{\mu\nu}(x)$ at $t=-t_s$. In the spatial direction, the finite-volume effects 
    are corrected by calculating $\delta_\text{IVR}^{(1)}$ and
    $\delta_\text{IVR}^{(2)}$ via the ground-state dominance.}
\end{figure}

Combining the corrections to the temporal truncation effects and finite-volume effects,
the main idea of IVR method is summarized in Fig.~\ref{IVR}.

\begin{figure}
        \centering
        \includegraphics[width=0.8\textwidth]{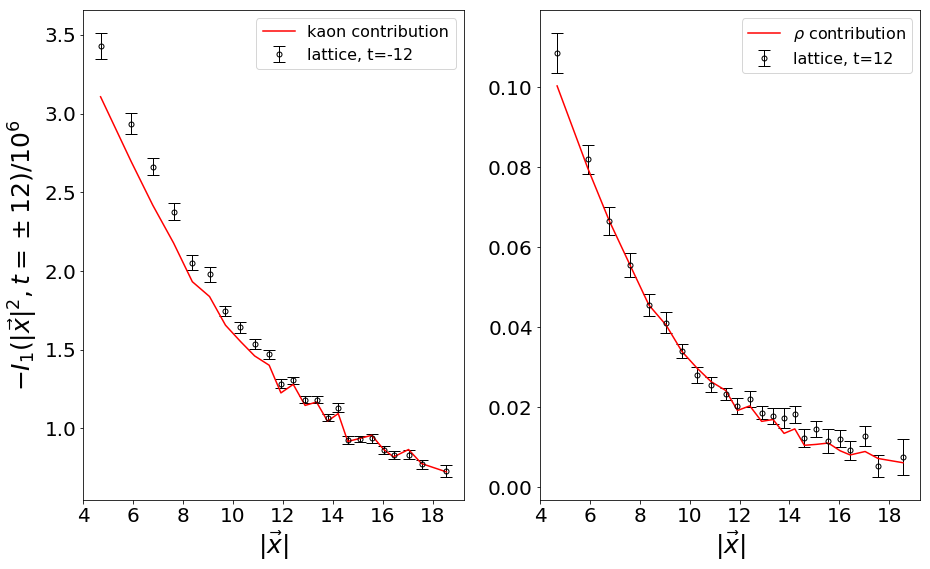}
        \caption{The comparison between lattice data $I_1(|\vec{x}|^2,t)=\delta^{\mu\nu} H^{\mu\nu}(x)$ and ground-state contribution. 
At large $|\vec{x}|$, the lattice data at $t=-12$ and 12 are well dominated by the kaon and rho state, respectively.}
\label{model}
\end{figure}

For sufficiently large time separation, $H^{(L),\mu\nu}(x)$ is saturated by ground-state contribution:
\begin{equation}
\label{Hmodel}
\begin{aligned}
H^{(L),\mu\nu}(\vec{x},t<-t_s)&=H_K^{(L),\mu\nu}(\vec{x},t)=\frac{1}{L^3}\sum_{\vec{p}}H_K^{\mu\nu}(\vec{p},t)e^{i\vec{p}\cdot\vec{x}}\\
H^{(L),\mu\nu}(\vec{x},t>t_s')&=H_\rho^{(L),\mu\nu}(\vec{x},t)=\frac{1}{L^3}\sum_{\vec{p}}H_\rho^{\mu\nu}(\vec{p},t)e^{i\vec{p}\cdot\vec{x}}.
\end{aligned}
\end{equation}
Here the hadronic kernels $H_{K,\rho}^{\mu\nu}(\vec{p},t)$ can be written in terms of form factors, whose explicit forms are
determined from lattice data $H^{(L),\mu\nu}(\vec{x},t)$. For more detailed discussions, we refer to Appendix~\ref{sect:form_factor}.
In Fig.~\ref{model}, we show the scalar function $I_1(|\vec{x}|^2,t)=\delta^{\mu\nu} H^{\mu\nu}(x)$ at $t=\pm12$ as an example that 
the lattice data are well dominated by the kaon and rho state.
The consistency between lattice data and ground-state contribution at long distance has also been checked for other scalar functions.
As a next step, we reconstruct the infinite-volume hadronic function $H_{K,\rho}^{\mu\nu}(x)$ 
and calculate correction $\delta_\text{IVR}$ through
\begin{equation}
\label{dIVR2}
\begin{aligned}
\delta_{\text{IVR}} &\approx \delta_{\text{IVR},K}+\delta_{\text{IVR},\rho}\\
\delta_{\text{IVR},K}&=
\int_V d^4 x\, e^{Et-i\vec{p}\cdot\vec{x}}(H_K^{\mu\nu}(x)-H_K^{(L),\mu\nu}(x))
+\int_{>V}d^4 x\,e^{Et-i\vec{p}\cdot\vec{x}}H_K^{\mu\nu}(x)\\
\delta_{\text{IVR},\rho}&=
\int_V d^4 x\, e^{Et-i\vec{p}\cdot\vec{x}}(H_\rho^{\mu\nu}(x)-H_\rho^{(L),\mu\nu}(x))
+\int_{>V}d^4 x\,e^{Et-i\vec{p}\cdot\vec{x}}H_\rho^{\mu\nu}(x).
\end{aligned}
\end{equation}
Using the approach described above, we can also apply the finite-volume correction to the scalar functions and obtain
\begin{equation}
\delta^\text{IVR}_i(L)\approx \delta^\text{IVR}_{i,K}(L)+\delta^\text{IVR}_{i,\rho}(L)
\end{equation}
with
\begin{equation}
\begin{aligned}
\delta^\text{IVR}_{i,K/\rho}(L)=\tilde{I}^\text{IVR}_{i,K/\rho}(\rho_1,\rho_2)-\tilde{I}_{i,K/\rho}^\text{IVR}(\rho_1,\rho_2;L)
\end{aligned}
\end{equation}
where the subscript $K$ or $\rho$ is used to indicate the 
scalar functions compiled from the ground-state contribution.
A parameter $L$ is introduced to
specify the scalar functions in the finite volume. 
For scalar functions in the
infinite volume, $\tilde{I}_{i,K/\rho}^\text{IVR}(\rho_1,\rho_2)$, it can be approximated
by $\tilde{I}_{i,K/\rho}^\text{IVR}(\rho_1,\rho_2;L_\infty)$ with a sufficiently large
$L_\infty$.

\subsection{\label{section3.4}Computation of the decay width}
\begin{figure}
	\centering
	\includegraphics[width=0.7\textwidth]{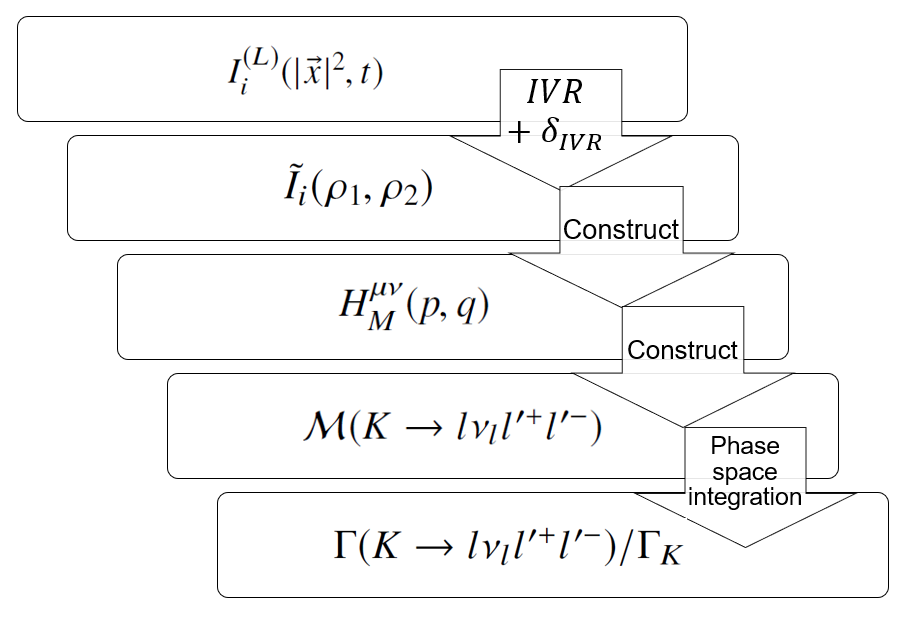}
	\caption{Lattice calculation procedures of the decay width.\label{fig:Calc}}
\end{figure}

In this section, we summarize the procedure to compute the decay width of
$K\to\ell\nu\ell'^+\ell'^-$. The outline of the main steps is shown in
Fig.~\ref{fig:Calc}.

In our calculation, the kaon mass $m_{K,\mathrm{lat}}=0.5057(13)$ GeV is slightly
larger than its physical value $m_{K,\mathrm{phy}}=0.493677(16)$ GeV~\cite{Tanabashi:2018oca}. It affects the decay width in two ways: 1) the $m_K$ dependence of hadronic function and 2) the change of phase space.
In order to reduce the latter,
for any dimensional
quantities $O^{[n]}$ with dimension $n$, we rescale them as
\begin{equation}
    \bar{O}^{[n]}=O^{[n]}\xi_K^n,\quad \xi_K\equiv
    \frac{m_{K,\mathrm{phy}}}{m_{K,\mathrm{lat}}}.
\end{equation}
For example, the decay constant and the hadronic function are rescaled as
\begin{equation}
\bar{f}_K=\xi_Kf_K,\quad
    \bar{H}_M^{\mu\nu}(p,q)=\xi_K H^{\mu\nu}_M(p,q).
\end{equation}
and the decay amplitudes originally defined in Eq.~(\ref{Msub}) are rescaled as
\begin{equation}
    \label{Mcalc}
    \bar{\mathcal{M}}_{D}=\xi_K\mathcal{M}_{D},\quad
    \bar{\mathcal{M}}_{E}=\xi_K\mathcal{M}_{E}.
\end{equation}
As the Fermi constant $G_F$ is a fixed coefficient,
$\mathcal{M}_{D,E}$ are considered here as the dimension-1 quantities. If there exists a phase-space integral, then $\frac{O^{[n]}}{m_{K,\mathrm{lat}}^n}$ relies on the dimensionless variables $\frac{P^2}{m_{K,\mathrm{lat}}^2}$ and
$\frac{P\cdot Q}{m_{K,\mathrm{lat}}^2}$. These variables take the same integral range as the ones in the physical case. We then multiply $\frac{O^{[n]}}{m_{K,\mathrm{lat}}^n}$ by a factor of $m_{K,\mathrm{phy}}^n$ to obtain a dimensional quantity.

For the lattice calculation of $\operatorname{Br}[K\to \ell\nu_\ell
\ell'^+\ell'^-]$ with $\ell= \ell'$, 
we use the Monte-Carlo integration. Within the allowed phase-space range, the
five parameters, $(x_{12},x_{34},y_{12},y_{34},\phi)$, are randomly generated $N_{MC}$ times. 
Given each momentum setup, $\bar{H}_{M}^{\mu\nu}(p_{12},q)$ and
$\bar{H}_{M}^{\mu\nu}(p_{14},q)$ are calculated using the IVR method. 
In order to get the decay amplitude in Eq.~(\ref{Mcalc}), numerical realization
of the spinor products are utilized. 
The branching ratio is calculated as follows
\begin{eqnarray}
    \operatorname{Br}[K\to\ell\nu_\ell\ell^+\ell^-]&&=\frac{1}{2 m_K \Gamma_{K}}
    \int d
    \Phi_{4}\,\left(\left|\bar{\mathcal{M}}_D\right|^{2}+\left|\bar{\mathcal{M}}_E\right|^{2}+2\operatorname{Re}[\bar{\mathcal{M}}_D\bar{\mathcal{M}}^*_E]\right)\nonumber\\
    &&=\frac{1}{2 m_K \Gamma_{K}}
    \frac{S}{N_{MC}}\sum_{i=1}^{N_{MC}}\frac{\mathcal{S}\lambda
    m_K^4}{2^{14}\pi^6}\left(\left|\bar{\mathcal{M}}_D\right|^{2}+\left|\bar{\mathcal{M}}_E\right|^{2}+2\operatorname{Re}[\bar{\mathcal{M}}_D\bar{\mathcal{M}}^*_E]\right)_i
\end{eqnarray}
where $S$ is the hypervolume of the integration range and $m_K=m_{K,\mathrm{phy}}$.

In the practical calculation, we choose $N_{MC}= 10000$, and confirm that the Monte-Carlo
error is much less than the statistical error.
Considering the fact that the construction of the scalar functions is nontrivial, the
Monte-Carlo integration thus provides an easily implemented approach to determine the decay
width.

\section{\label{section4}Numerical results}
\subsection{\label{section4.1}Lattice setup}
In the lattice calculation, we use a gauge ensemble with $N_f=2+1+1$-flavor twisted mass fermion generated by
ETM Collaboration~\cite{Alexandrou:2018egz}. The light quark mass is
unphysical with $m_\pi=352$ MeV. We tune the valence strange quark mass to have the kaon mass close
to its physical value. 
Parameters of the gauge
ensemble are listed in Table~\ref{ens} together with the information of $\Delta
T$, whose value shall be set sufficiently large to suppress the excited-state
contamination.
\begin{table}
	\begin{ruledtabular}
	\begin{tabular}{ccccccc}
		Label&$L^3\times T$&$a^{-1}$& $N_{conf}$&$m_\pi$ &$m_K$ &$\Delta T$\\
		\hline
		cA211b.53.24&$24^3\times 48$& 2.12 GeV & 51& 0.3515(15) GeV &0.5057(13) GeV&10
	\end{tabular}
	\end{ruledtabular}
    \caption{\label{ens}Information of lattice setup.}
\end{table}

In order to calculate hadronic functions, three-point correlation functions are
calculated on lattice. The initial kaon state is created using Coulomb
gauge-fixed wall source operator. 
We place two point-source propagators and one wall-source propagator at each time slice 
and perform a time translation average over all time slices to obtain the three-point function.

\subsection{\label{section4.2}Results}
\begin{figure}[htbp]
	\centering
	\includegraphics[width=0.9\textwidth]{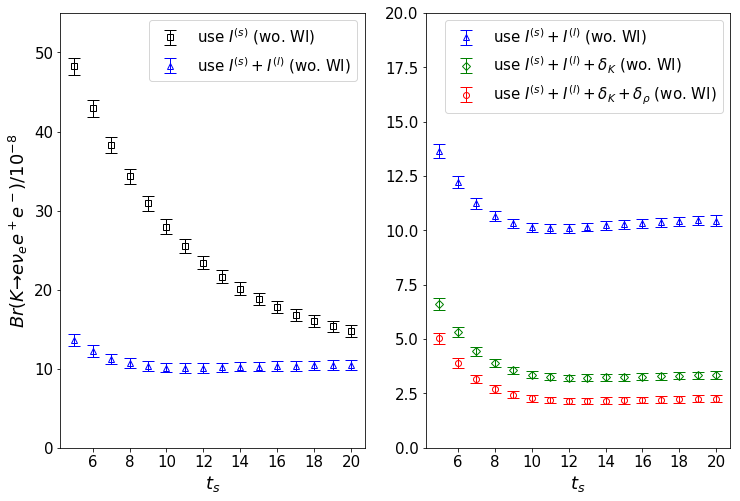}
	\caption{IVR results for $K\to e\nu_e e^+ e^-$
    ($L_\infty=72$). The constraint from Ward identity is not used here. In the left-hand panel, the black-square data points are compiled using
the short-distance contribution $\tilde{I}_i^{(s)}$ while the blue-triangle ones using both short-distance and long-distance contributions, namely $\tilde{I}_i^{(s)}$+$\tilde{I}_i^{(l)}$. In the right-hand panel, it shows that the results shift due to the corrections from the kaon state, $\delta^{\text{IVR}}_{i,K}$ and the rho state, $\delta^{\text{IVR}}_{i,\rho}$.}
\label{fig:IVR_wo_WI}
\end{figure}

\begin{figure}[htbp]
        \centering
        \includegraphics[width=0.9\textwidth]{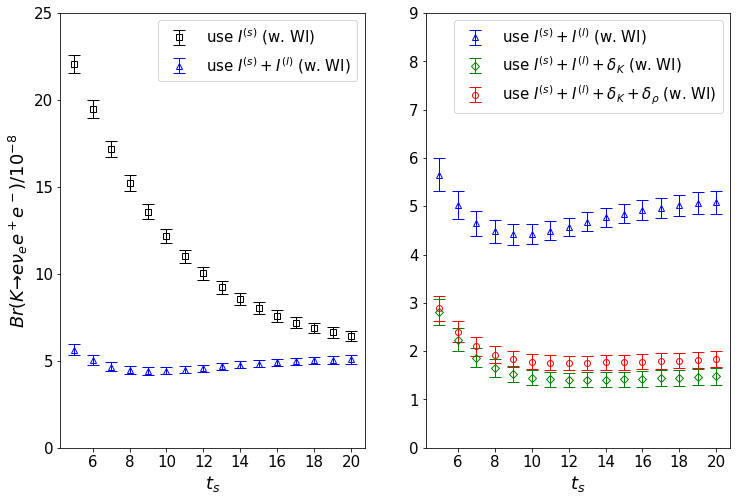}
        \caption{Similar as Fig.~\ref{fig:IVR_wo_WI}, but utilizing the constraint from Ward identity.}
\label{fig:IVR_w_WI}
\end{figure}

In Figs.~\ref{fig:IVR_wo_WI} and \ref{fig:IVR_w_WI}, we take $K\to e\nu_e e^+ e^-$ as an example
to show the results of branching ratio as a function of $t_s$. 
In the left-hand panel, 
the black-square data points are compiled using
the short-distance contribution $\tilde{I}_i^{(s)}$ while the blue-triangle ones using both short-distance and long-distance contributions.
A significant temporal truncation effect is found, demonstrating the
importance of the IVR correction. The time $t_s$ needs to be sufficiently large
to guarantee the ground-intermediate-state dominance. 
The figure shows that 
starting from $t_s\approx12$, the branching ratio is
independent from the choice of $t_s$.

In the right-hand panel of Figs.~\ref{fig:IVR_wo_WI} and \ref{fig:IVR_w_WI}, the effects of $\delta^{\text{IVR}}_{i,K}$ and $\delta^{\text{IVR}}_{i,\rho}$ are shown.
These corrections are made with the
choice of $L_\infty=72$.
We find a large correction from the kaon state and a relatively smaller correction from the rho state. Given the $L=2.2$ fm lattice used in our calculation, 
it is essential to include both corrections.

\begin{figure}
        \centering
        \includegraphics[width=0.9\textwidth]{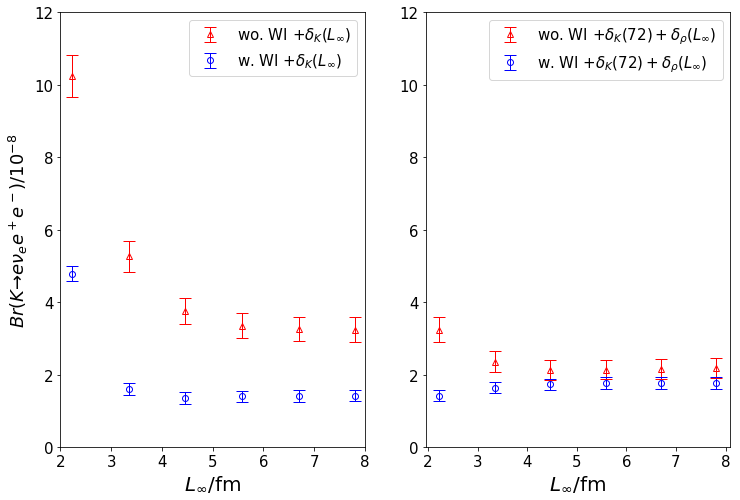}
        \caption{\label{Linf}
        Examination of $L_{\infty}$-dependence in the $\delta_i^{\text{IVR}}$ corrections.
        Here we use $K\to e\nu_e e^+ e^-$ decay as an example. In the left-hand panel we show the finite-volume correction $\delta^{\text{IVR}}_{i,K}$ as a function of $L_\infty$. In the right-hand panel, we fix
  $\delta^{\text{IVR}}_{i,K}$ at $L_\infty=72$ and show the correction $\delta^{\text{IVR}}_{i,\rho}$ as a function of $L_\infty$.
        When $L_\infty=L=24$ (corresponding to $2.2$ fm), no correction is made and $\delta_{\text{IVR}}=0$.
        By increasing $L_\infty$, $\delta^{\text{IVR}}_i$ exponentially converges and
        $L_\infty=72$ (corresponding to $6.7$ fm) is a sufficiently large lattice size to approximate the
        infinite one.}
\end{figure}

For the results shown in Figs.~\ref{fig:IVR_wo_WI} and \ref{fig:IVR_w_WI}, the parameter $L_\infty$ is chosen
as $L_\infty=72$. In Fig.~\ref{Linf} we examine the $L_\infty$-dependence
of the results. In the left-hand panel, it shows that 
the corrections of $\delta^{\text{IVR}}_{i,K}$
become much smaller when using the constraint from Ward identity.
In the right-hand panel, it shows that by including the
corrections of $\delta^{\text{IVR}}_{i,\rho}$, the lattice results using or not using Ward identity constraint start to converge at large $L_\infty$.
 We also confirm that as $L_\infty$ increases, the finite-volume
effects exponentially suppress and $L_\infty=72$ is an appropriate choice to
approximate the infinitely large spatial extent.

\begin{figure}
        \centering
        \includegraphics[width=1\textwidth]{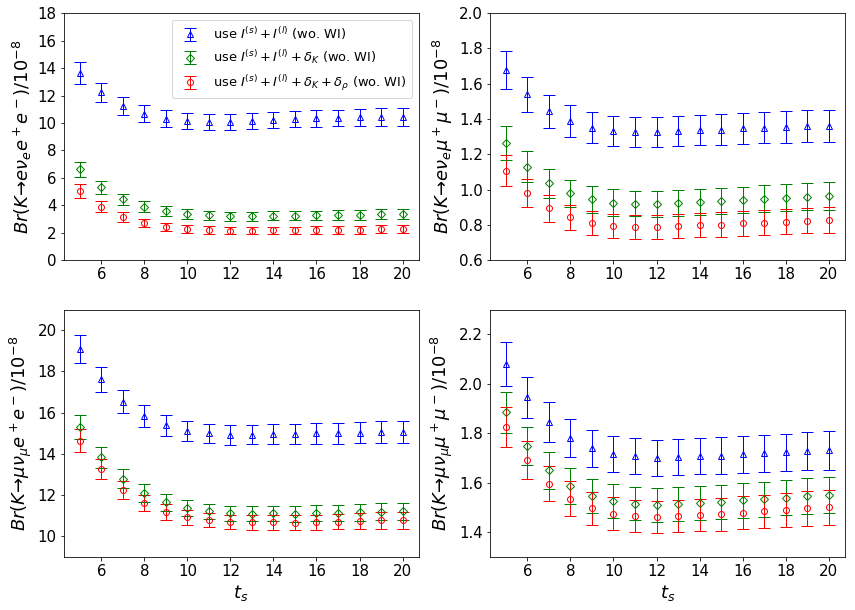}
        \caption{
        IVR results for four channels of $K\to \ell\nu_\ell \ell'^+ \ell'^-$. We don't utilize the Ward-identity constraint here. The upper-left figure for $K\to e\nu_e e^+ e^-$ has been shown in the right-hand
    panel of Fig.~\ref{fig:IVR_wo_WI}. We put it here for the sake of an easier comparison with other three channels. Through the comparison we find that the $\delta^{\text{IVR}}_{i,K}$ corrections are important for all the channels. The corrections from the $\delta^{\text{IVR}}_{i,\rho}$ are less significant but still comparable to the size of the statistical errors.}
\label{fig:resall_S1}
\end{figure}

\begin{figure}
        \centering
        \includegraphics[width=1\textwidth]{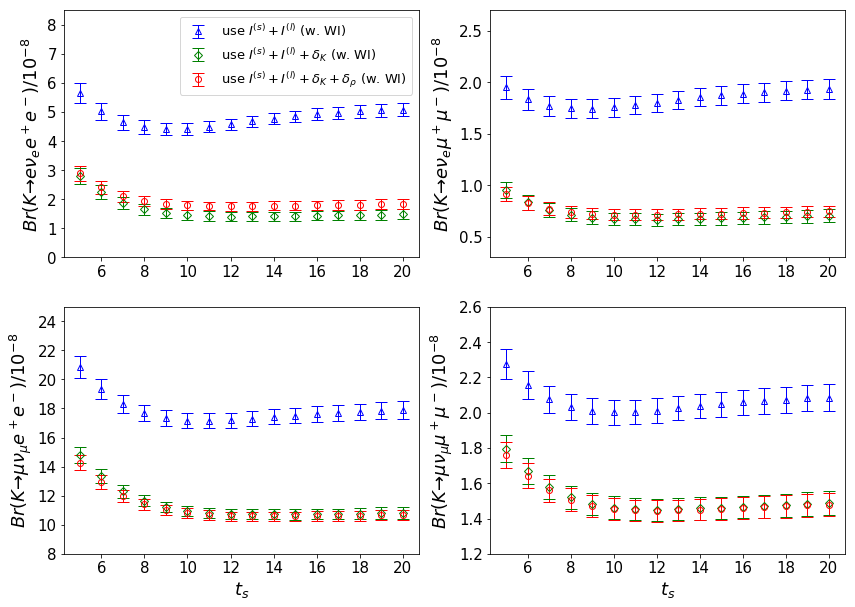}
        \caption{Similar as Fig.~\ref{fig:resall_S1}, but utilizing the constraint from Ward identity.}
\label{fig:resall_S2}
\end{figure}

\begin{table*}
        \begin{ruledtabular}
        \begin{tabular}{ccc}
                Channel & using Ward-identity constraint & not using Ward-identity constraint \\
                \hline
        $\operatorname{Br}[K\to e\nu_e e^+e^-]$ & $1.77(16)\times10^{-8}$ & $2.18(27)\times10^{-8}$ \\
        $\operatorname{Br}[K\to \mu\nu_\mu e^+e^-]$ & $10.59(33)\times10^{-8}$ & $10.68(36)\times10^{-8}$ \\
	$\operatorname{Br}[K\to e\nu_e \mu^+\mu^-]$ & $0.72(5)\times10^{-8}$ & $0.80(7)\times10^{-8}$ \\
	$\operatorname{Br}[K\to \mu\nu_\mu \mu^+\mu^-]$ & $1.45(6)\times10^{-8}$ & $1.47(7)\times10^{-8}$ \\
        \end{tabular}
\end{ruledtabular}
    \caption{A comparison of the lattice results by using or not using the constraint from Ward identity.}
\label{tab:results_w_or_wo_WI}
\end{table*}

In Figs.~\ref{fig:resall_S1} and \ref{fig:resall_S2}, IVR results for all the four channels of $K\to \ell\nu_\ell \ell'^+ \ell'^-$ are shown. 
The final results of branching ratio are summarized in Table~\ref{tab:results_w_or_wo_WI}, where we make a comparison between using or not using the constraint from the Ward identity. 
The results are very consistent.
As the former
suffers from much less finite-volume effects in the $K\to e\nu_e e^+ e^-$ decay, we quote the corresponding results in Table~\ref{resintro} for a comparison with ChPT and experiments. All the lattice results
are compiled using the fitting range $t_s\in [12,17]$ and including the corrections from both $\delta^{\text{IVR}}_{i,K}$ and $\delta^{\text{IVR}}_{i,\rho}$.
We find that the lattice results are comparable to experimental or ChPT ones. The systematic
errors of our results mainly come from unphysical quark masses, lattice
artifacts and residual finite volume effects, which are left for a future study.

\section{\label{section5}Conclusion}
In this work, we build a lattice calculation procedure to determine $K\to
\ell\nu_\ell \ell'^+\ell'^-$ decay width by solving a series of technical
problems.
IVR method is used to reduce temporal truncation effects and finite-volume effects. 
Other approaches, such as scalar function method and Monte-Carlo phase-space integration, are proposed to 
simplify the calculation. Using these techniques, a practical methodology is
developed to
compute decay width with four daughter particles in the final state, as summarized in Fig.~\ref{fig:Calc}.

Using this methodology, we perform a realistic lattice calculation of $K\to
\ell\nu_\ell \ell'^+\ell'^-$ decay width using an ensemble with pion mass $352$
MeV and kaon mass $506$ MeV, and obtain the branching ratios comparable to ChPT or experimental results. 
Through the calculation, we demonstrate the capability of lattice QCD to improve
Standard Model prediction in $K\to \ell\nu_\ell \ell'^+ \ell'^-$ decay width. 
By examining the $t_s$-dependence and $L_\infty$-dependence of decay width, we
show that the IVR method is a vital approach to reduce the systematic effects.
Future work is still required to address the power-law finite-volume effects in
the subprocess of $K\to\pi\pi \ell\nu_\ell\to \ell\nu_\ell \ell'^+ \ell'^-$ and
make a full control of various systematic effects.

\begin{acknowledgments}

We thank Guido Martinelli for giving an inspiring lecture at the lattice QCD
    summer school held at Peking University on 2019, which motivates this work.
We thank ETM Collaboration for sharing the gauge configurations with us.
X.F. and L.C.J. gratefully acknowledge many helpful discussions with our colleagues from the
RBC-UKQCD Collaboration. X.F., X.Y.T. and T.W. are supported in part by NSFC of China under Grants No.
12125501, No. 12070131001, No. 12141501, and No. 11775002
and National Key Research and Development
    Program of China under Contracts No. 2020YFA0406400. L.C.J. acknowledges support by DOE Office of Science Early Career Award DE-SC0021147 and DOE grant DE-SC0010339.
The calculation was carried out on TianHe-3 (prototype) at Chinese National Supercomputer Center in Tianjin.

\end{acknowledgments}

\appendix

\section{Formulae in scalar function method}
\label{B}
\subsection{Infinite-volume case}

In Sec.~\ref{section3}, the scalar function method is described by Eqs.~(\ref{Scalarmethod1}) and (\ref{Scalarmethod2}). 
In this section, we will give the detailed expression to calculate
$\tilde{I}_i(\rho_1,\rho_2)$ and then discuss the approach to construct $H^{\mu\nu}(P,Q)$
using the scalar functions.

We start with the relation
\begin{equation}
    \label{eq:mom_average}
    H^{\mu\nu}(P,Q)=-i\int d^4x\,
    e^{Et-i\vec{p}\cdot\vec{x}}H^{\mu\nu}(x).
\end{equation}
Combining Eq.~(\ref{eq:mom_average}) with the Lorentz factors given in
Eq.~(\ref{eq:scalar_mom}), we can obtain the relation between
$\tilde{I}_i(\rho_1,\rho_2)$ and $I_i(|\vec{x}|^2,t)$. Here we give the detailed
expressions of $\tilde{I}_i(\rho_1,\rho_2)$ as
\begin{subequations}
\label{I1}
\begin{equation}
    \tilde{I}_{1}(\rho_1,\rho_2)=im_K^2\int d^4
    x\,e^{Et}j_0(\varphi)I_1(|\vec{x}|^2,t),
\end{equation}
\begin{equation}
    \tilde{I}_{2}(\rho_1,\rho_2)=im_K^2\int d^4
    x\,e^{Et}j_0(\varphi)I_2(|\vec{x}|^2,t),
\end{equation}
\begin{equation}
    \tilde{I}_{3}(\rho_1,\rho_2)=im_KE\int d^4 x\,e^{Et}j_0(\varphi)I_2(|\vec{x}|^2,t)
    -im_K|\vec{p}|\int d^4
    x\,e^{Et}\frac{j_1(\varphi)}{|\vec{x}|}
    I_3(|\vec{x}|^2,t),
\end{equation}
\begin{equation}
    \tilde{I}_{4}(\rho_1,\rho_2)=im_KE\int d^4 x\,e^{Et}j_0(\varphi)I_2(|\vec{x}|^2,t)
    -im_K|\vec{p}|\int d^4
    x\,e^{Et}\frac{j_1(\varphi)}{|\vec{x}|}
    I_4(|\vec{x}|^2,t),
\end{equation}
\begin{eqnarray}
\label{I5}
    \tilde{I}_{5}(\rho_1,\rho_2)=&&iE^2\int d^4
    x\,e^{Et}j_0(\varphi)I_2(|\vec{x}|^2,t)
    +i|\vec{p}|^2\int d^4
    x\,e^{Et}\frac{j_2(\varphi)}{|\vec{x}|^2}
    I_5(|\vec{x}|^2,t)
    \\
    &&-i|\vec{p}|\int d^4
    x\,e^{Et}\frac{j_1(\varphi)}{|\vec{x}|}[EI_3(|\vec{x}|^2,t)+EI_4(|\vec{x}|^2,t)+I_1(|\vec{x}|^2,t)-I_2(|\vec{x}|^2,t)]\nonumber
\end{eqnarray}
\begin{equation}
\label{I6}
    \tilde{I}_{6}(\rho_1,\rho_2)=-im_K|\vec{p}|\int d^4
    x\,e^{Et}\frac{j_1(\varphi)}{|\vec{x}|}I_6(|\vec{x}|^2,t).
\end{equation}
\end{subequations}
Note that in the continuum theory, the scalar functions $\tilde{I}_{i}(\rho_1,\rho_2)$ do not
depend on the direction of $\vec{p}$. Thus, in the derivation of the above
equations we have performed an average over the solid angle of $\vec{p}$. 
After the average, the factor $e^{-i\vec{p}\cdot\vec{x}}$ is converted into a
spherical Bessel function $j_0(\varphi)$, with $\varphi=|\vec{p}||\vec{x}|$. 
In total, three spherical Bessel functions appear in
Eq.~(\ref{I1}). They take the standard definition as
\begin{equation}
    j_0(\varphi)\equiv \frac{\sin{\varphi}}{\varphi},\quad
    j_1(\varphi)\equiv\frac{\sin\varphi-\varphi\cos\varphi}{\varphi^2},\quad
    j_2(\varphi)\equiv\frac{(3-\varphi^2)\sin\varphi-3\varphi\cos\varphi}{\varphi^3}.
\end{equation}
In the numerical calculation, when the variables
$\rho_1$ and $\rho_2$ are given, the values of $|\vec{p}|$ and $E$ can be
determined through
\begin{equation}
    |\vec{p}|=\frac{1}{2}m_K\sqrt{(1+\rho_1-\rho_2)^2-4\rho_1},\quad
    E=\frac{1}{2}m_K(1+\rho_1-\rho_2).
\end{equation}

Once these scalar functions $\tilde{I}_i(\rho_1,\rho_2)$ are available,
$H^{\mu\nu}(P,Q)$ can be easily constructed using Eq.~(\ref{Scalarmethod1}). In
numerical calculation, $w_i(P,Q)$ in Eq.~(\ref{Scalarmethod1}) is implicitly
derived using following procedures:

\begin{enumerate}
	\item A general factorization of $H^{\mu\nu}(P,Q)$ is used with
	\begin{eqnarray}
	\label{fac}
	H^{\mu \nu}(P, Q)&&=a\left(\rho_{1}, \rho_{2}\right) P^{\mu}
        Q^{\nu}+b\left(\rho_{1}, \rho_{2}\right) P^{\nu}
        Q^{\mu}+c\left(\rho_{1}, \rho_{2}\right) P^{\mu}
        P^{\nu}\nonumber\\&&+d\left(\rho_{1}, \rho_{2}\right) Q^{\mu}
        Q^{\nu}+e\left(\rho_{1}, \rho_{2}\right) \delta^{\mu \nu}
        m_{K}^{2}+f\left(\rho_{1}, \rho_{2}\right) \varepsilon^{\mu \nu \alpha
        \beta} P^{\alpha} Q^{\beta}.
    \end{eqnarray}
	
	\item $\tilde{I}_{i}(\rho_1,\rho_2)(i=1\dots 6)$ and
        $a(\rho_1,\rho_2),\dots,f(\rho_1,\rho_2)$ are related by a simple linear
        transformation. We can then solve the solution for $a(\rho_1,\rho_2),\dots,f(\rho_1,\rho_2)$ and construct $H^{\mu\nu}(P,Q)$ using Eq.~(\ref{fac}). 
\end{enumerate}

\subsection{Scalar functions with the IVR corrections}
In Sec.~\ref{section3}, IVR method is proposed to correct 
the temporal truncation effects and
the finite volume effects for the hadronic functions $H^{\mu\nu}(P,Q)$.
In this section, we show how to apply the IVR method to the scalar
functions.

We shall point out first that the calculation of $\tilde{I}_6(\rho_1,\rho_2)$ does not require the IVR correction.
It is because
$\tilde{I}_6(\rho_1,\rho_2)$ is projected out by using the Lorentz factor of
$\varepsilon_{\mu\nu\alpha\beta}P^{\alpha}Q^\beta$.
In this quantity, the intermediate states are given by the states heavier than initial kaon
state and thus the temporal truncation effects and finite-volume effects can be
neglected.

In the calculation of $\tilde{I}_i(\rho_1,\rho_2)$, with $i=1,\cdots,5$, we also
use $t=-t_s$ to separate the time integral into the short-distance part and the
long-distance part
\begin{equation}
	\tilde{I}_i(\rho_1,\rho_2)=\tilde{I}^{(s)}_i(\rho_1,\rho_2)+\tilde{I}^{(l)}_i(\rho_1,\rho_2).
\end{equation} 
For the short-distance part, we replace $\tilde{I}^{(s)}_i(\rho_1,\rho_2)$ by
the lattice data $\tilde{I}^{(s)}_i(\rho_1,\rho_2,L)$.
For the long-distance part, we use the lattice data of $I_i^{(L)}(|\vec{x}|^2,t)$ at
$t=-t_s$ as input. Through the kaon-intermediate-state dominance,
$\tilde{I}^{(l)}_i(\rho_1,\rho_2,L)$ can be reconstructed. 
The detailed expressions are given as
\begin{subequations}
	\label{IVRf}
	\begin{equation}
        \tilde{I}^{(l)}_{1}(\rho_1,\rho_2;L)=\frac{im_K^2}{E+E_K-m_K}\int d^3
        \vec{x}\, e^{-Et_s}j_0(\varphi)I^{(L)}_{1}(|\vec{x}|^2,t_s),
    \end{equation}
	
	\begin{equation}
        \tilde{I}^{(l)}_{2}(\rho_1,\rho_2;L)=\frac{im_K^2}{E+E_K-m_K}\int d^3
        \vec{x}\, e^{-Et_s}j_0(\varphi)I^{(L)}_{2}(|\vec{x}|^2,t_s),
	\end{equation}
	
	\begin{eqnarray}
	\tilde{I}^{(l)}_{3}(\rho_1,\rho_2;L)=&&\frac{im_K}{E+E_K-m_K}\left[E\int d^3
        \vec{x}\, e^{-Et_s}j_0(\varphi)I^{(L)}_{2}(|\vec{x}|^2,t_s)\right.\\
        &&
        \hspace{2.5cm}\left.+|\vec{p}|\int d^3 \vec{x}\,
        e^{-Et_s}\frac{j_1(\varphi)}{|\vec{x}|}
        I^{(L)}_{3}(|\vec{x}|^2,t_s)\right],\nonumber
	\end{eqnarray}	
	
	\begin{eqnarray}
	\tilde{I}^{(l)}_{4}(\rho_1,\rho_2;L)=&&\frac{im_K}{E+E_K-m_K}\left[E\int d^3
        \vec{x}\, e^{-Et_s}j_0(\varphi)I^{(L)}_{2}(|\vec{x}|^2,t_s)\right.\\
        &&\hspace{2.5cm}\left.+|\vec{p}|\int d^3 \vec{x}\,
        e^{-Et_s}\frac{j_1(\varphi)}{|\vec{x}|}
        I^{(L)}_{4}(|\vec{x}|^2,t_s)\right],\nonumber
	\end{eqnarray}
	
	\begin{eqnarray}
	\tilde{I}^{(l)}_{5}(\rho_1,\rho_2;L)=&&\frac{im_K}{M(E+E_K-m_K)}\left[E^2\int
        d^3 \vec{x}\,e^{-Et_s}j_0(\varphi)I^{(L)}_{2}(|\vec{x}|^2,t_s)\right.\nonumber\\
        &&\hspace{3.5cm}+|\vec{p}|\int d^3
        \vec{x}\,e^{-Et_s}\frac{j_1(\varphi)}{|\vec{x}|}\\
        &&\hspace{1.5cm}\times[EI^{(L)}_{3}(|\vec{x}|^2,t_s)+EI^{(L)}_{4}(|\vec{x}|^2,t_s)+I^{(L)}_{1}(|\vec{x}|^2,t_s)-I^{(L)}_{2}(|\vec{x}|^2,t_s)]\nonumber\\
        &&\hspace{3.5cm}\left.-|\vec{p}|^2\int d^3 \vec{x}\,
        e^{-Et_s}\frac{j_2(\varphi)}{|\vec{x}|^2}
        I^{(L)}_{5}(|\vec{x}|^2,t_s)\right].\nonumber
	\end{eqnarray}	
\end{subequations}
The scalar functions calculated through the IVR method are given by
\begin{equation}
    \tilde{I}_i^\text{IVR}(\rho_1,\rho_2;L)=\tilde{I}^{(s)}_i(\rho_1,\rho_2;L)+\tilde{I}^{(l)}_i(\rho_1,\rho_2;L),\quad
    \mbox{for }i=1,\cdots,5.
\end{equation}

As a next step, we perform the finite-volume correction by introducing
$\delta_i^\text{IVR}(L)$ for each scalar function
\begin{equation}
\tilde{I}_i(\rho_1,\rho_2)
    =\tilde{I}_i^\text{IVR}(\rho_1,\rho_2;L)+\delta_i^\text{IVR}(L).
\end{equation} 
Here $\delta_i^\text{IVR}(L)$ can be
approximated by the kaon- and rho- state contribution
\begin{equation}
 \delta_i^\text{IVR}(L)\approx \tilde{I}_{i,K}^\text{IVR}(\rho_1,\rho_2)-\tilde{I}_{i,K}^\text{IVR}(\rho_1,\rho_2;L)
\tilde{I}_{i,\rho}^\text{IVR}(\rho_1,\rho_2)-\tilde{I}_{i,\rho}^\text{IVR}(\rho_1,\rho_2;L).
\end{equation}
In practice, $\tilde{I}_{i,K}^\text{IVR}(\rho_1,\rho_2)$ and $\tilde{I}_{i,\rho}^\text{IVR}(\rho_1,\rho_2)$ in the infinite volume can
be replaced by $\tilde{I}_{i,K}^\text{IVR}(\rho_1,\rho_2;L_\infty)$ and $\tilde{I}_{i,\rho}^\text{IVR}(\rho_1,\rho_2;L_\infty)$ with $L_\infty\gg
L$.

\section{Estimate the finite-volume correction from the kaon and rho states}

\label{sect:form_factor}

In Eq.~(\ref{Hmodel}) the hadronic kernels are defined as:
\begin{equation}
\begin{aligned}
\label{Hmodelp}
H^{\mu\nu}_K(\vec{p},t)&=\frac{e^{(E_K-m_K)t}}{2E_K}\langle
    0|J_W^\nu(0)|K(p_K)\rangle\langle
    K(p_K)|J_{\text{em}}^\mu(0)|K\rangle,\\
H^{\mu\nu}_\rho(\vec{p},t)&=\frac{e^{-E_\rho t}}{2E_\rho}\sum_\lambda\langle
    0|J_{\text{em}}^\mu(0)|\rho(p_\rho,\lambda)\rangle\langle
    \rho(p_\rho,\lambda)|J_W^\nu(0)|K\rangle,
\end{aligned}
\end{equation}
where $\lambda$ is the polarized direction for vector meson. 
$p_{K}=(iE_{K},-\vec{p})$ and $p_{\rho}=(iE_{\rho},\vec{p})$ are the four-momentum for intermediate states with the energies
$E_{K}=\sqrt{\vec{p}^2+m_{K}^2}$, $E_{\rho}=\sqrt{\vec{p}^2+m_{\rho}^2}$. 
We define the momentum transfer between initial kaon and intermediate particles as $q_K=Q-p_{K}$, and $q_\rho=Q-p_{\rho}$. 
The relevant hadronic matrix elements are given by
\begin{equation}
\begin{aligned}
	\langle 0|J_W^\mu(0)|K(p_K)\rangle=&f_K p_K^\mu,\\
	\langle 0|J_{\text{em}}^\mu(0)|\rho(p_\rho,\lambda)\rangle=&f_\rho \epsilon^\mu(p_\rho,\lambda),\\
	\langle K(p_K)|J_{\text{em}}^\mu(0)|K\rangle=&F^{(K)}(q_K^2)(Q+p_K)^\mu,\\
	\langle \rho(p_\rho,\lambda)|J_W^\mu(0)|K\rangle=&\frac{2V(q_\rho^2)}{m_K+m_\rho}\varepsilon^{\mu\nu\alpha\beta}\epsilon^\nu(p_\rho,\lambda)p_\rho^\alpha Q^\beta-(m_K+m_\rho)A_1(q_\rho^2)\epsilon^\mu(p_\rho,\lambda)\\
	&+\frac{A_2(q_\rho^2)(\epsilon\cdot Q)}{m_K+m_\rho}(Q+p_\rho)^\mu+\frac{2m_\rho A(q_\rho^2)(\epsilon\cdot Q)}{q^2_\rho}(Q-p_\rho)^\mu,
\end{aligned}
\end{equation}
where $f_K$ and $f_\rho$ are decay constants for $K$ and $\rho$. $\epsilon^\mu(p_\rho,\lambda)$ is the polarization vector. $F^{(K)}(q^2)$ is kaon electromagnetic form factor. $V(q^2)$, $A_1(q^2)$, $A_2(q^2)$ and $A(q^2)$ are the form factors for the semileptonic decays, with the convention adopted from Refs.~\cite{Richman:1995wm,Bowler:2004zb}. The form factor $A(q^2)$ 
approaches to zero at the limit of $q^2\to0$~\cite{Richman:1995wm}.

We use the simple parametrization for the above form factors
\begin{equation}
\label{eq:fit_form}
\begin{aligned}
&F^{(K)}(q^2)=\frac{1}{1+\frac{\langle r_K^2\rangle}{6}q^2},\quad V(q^2)=\frac{v}{1+v'\frac{q^2}{(m_\rho+m_K)^2}},\\
&A_1(q^2)=\frac{a_1}{1+a'_1\frac{q^2}{(m_\rho+m_K)^2}},\quad A_2(q^2)=\frac{a_2}{1+a'_2\frac{q^2}{(m_\rho+m_K)^2}},\\
&A(q^2)=\frac{a'\frac{q^2}{(m_\rho+m_K)^2}}{1+a''\frac{q^2}{(m_\rho+m_K)^2}},
\end{aligned}
\end{equation}
with $\langle r_K^2\rangle$ the square of kaon charge radius and $v$, $v'$, $a_1$, $a_1'$, $a_2$, $a_2'$, $a'$ and $a''$ the free parameters.

\begin{table*}
        \begin{ruledtabular}
        \begin{tabular}{cccccc}
                $\frac{L}{2\pi}\vec{p}$ & (0,0,0)& (0,0,1)& (0,1,1)& (1,1,1)& (0,0,2)\\
                \hline
        $q_K^2$ [GeV$^2$] &0 &0.055 &0.096 & 0.131& 0.161 \\
        $q_\rho^2$ [GeV$^2$] &-0.129 &0.036 &0.178 & 0.305& 0.421 \\
        \end{tabular}
\end{ruledtabular}
    \caption{Momenta used in the determination of the form factors. \label{Table_momenta}}
\end{table*}

\begin{figure}
        \centering
        \includegraphics[width=0.95\textwidth]{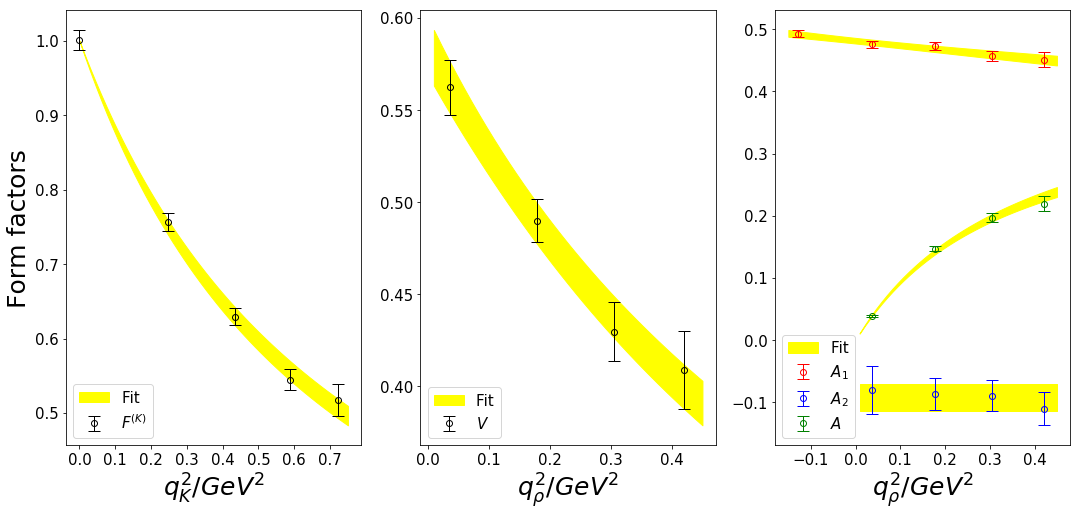}
        \caption{The lattice results of the form factors together with the fitting curve. \label{fig:form_factors}}
\end{figure}

\begin{table*}
        \begin{ruledtabular}
                \begin{tabular}{cccccccc}
                          $\langle r_K^2 \rangle$& $v$& $v'$& $a_1$& $a'_1$& $a_2$& $a'$ & $a''$\\
                        \hline
                         $0.32(2)$ fm$^2$ & $0.58(2)$ &$2.1(2)$ & $0.482(5)$& $0.30(7)$ &$-0.09(2)$ & $2.2(1)$ & $5.3(4)$\\
                \end{tabular}
        \end{ruledtabular}
        \caption{Determination of the parameters for the form factors. \label{table:form_factors}}
\end{table*}

We calculate the hadronic matrix elements at discrete lattice momenta listed in Table~\ref{Table_momenta}. Through the fit to the forms~(\ref{eq:fit_form}), we extract the parameters
shown in Table~\ref{table:form_factors}. The lattice data together with the fitting curves are plotted in Fig.~\ref{fig:form_factors}. Once $H^{\mu\nu}_{K/\rho}(\vec{p},t)$ are determined, we can estimate the finite-volume corrections $\delta^{\text{IVR}}_{i,K/\rho}$.
Note that the parametrization in Eq.~(\ref{eq:fit_form}) does bring in the model dependence, but only affects 
$\delta^{\text{IVR}}_{i,K/\rho}$. 
As far as the lattice size is sufficiently large, $\delta^{\text{IVR}}_{i,K/\rho}$ will vanish exponentially.

\bibliography{ref}
\end{document}